\documentclass{aa}  

\usepackage{graphicx}
\usepackage{txfonts}
\usepackage{graphicx}
\usepackage{lscape}
\usepackage{longtable}
\usepackage{natbib}
\usepackage{color}
\usepackage{array} 
\usepackage{array} 
\usepackage{tikz,array}
\usetikzlibrary{calc}
\usepackage{txfonts}
\usepackage{color}
\usepackage{multirow}
\usepackage{here}

\usepackage{txfonts}
\usepackage{amsmath,amstext}
\usepackage{graphicx}
\usepackage{epstopdf}
\usepackage{float}
\usepackage{array}
\usepackage{mathtools}
\usepackage{booktabs}
\usepackage{subfigure}
\usepackage{url}
\usepackage{helvet}
\usepackage{tabularx}
\usepackage{multirow}
\usepackage{natbib}
\usepackage[flushleft]{threeparttable}
\usepackage{lscape}
\usepackage{pdflscape}
\usepackage{longtable}
\usepackage{wasysym}
\usepackage{float}
\usepackage{relsize}
\usepackage{color}
\usepackage{breqn}
\usepackage{bm}

\usepackage[colorlinks, citecolor=blue, linkcolor=blue]{hyperref}
\hypersetup{colorlinks,breaklinks, linkcolor=blue,urlcolor=magenta, anchorcolor=blue,citecolor=blue}
\usepackage{scalerel}
\usepackage{tikz}
\usetikzlibrary{svg.path}
\definecolor{orcidlogocol}{HTML}{A6CE39}
\tikzset{
  orcidlogo/.pic={
    \fill[orcidlogocol] svg{M256,128c0,70.7-57.3,128-128,128C57.3,256,0,198.7,0,128C0,57.3,57.3,0,128,0C198.7,0,256,57.3,256,128z};
    \fill[white] svg{M86.3,186.2H70.9V79.1h15.4v48.4V186.2z}
                 svg{M108.9,79.1h41.6c39.6,0,57,28.3,57,53.6c0,27.5-21.5,53.6-56.8,53.6h-41.8V79.1z M124.3,172.4h24.5c34.9,0,42.9-26.5,42.9-39.7c0-21.5-13.7-39.7-43.7-39.7h-23.7V172.4z}
                 svg{M88.7,56.8c0,5.5-4.5,10.1-10.1,10.1c-5.6,0-10.1-4.6-10.1-10.1c0-5.6,4.5-10.1,10.1-10.1C84.2,46.7,88.7,51.3,88.7,56.8z};
  }
}
\newcommand\orcidicon[1]{\href{https://orcid.org/#1}{\mbox{\scalerel*{
\begin{tikzpicture}[yscale=-1,transform shape]
\pic{orcidlogo};
\end{tikzpicture}
}{|}}}}

\def\gcm3{\hbox{g cm$^{-3}$}}       %g.cm-3
\def\Mearth{\hbox{$\mathrm{M}_{\oplus}$}}
\def\Rearth{\hbox{$\mathrm{R}_{\oplus}$}}

\bibpunct{(}{)}{;}{a}{}{,} % to follow the A&A style

\newcommand\gO{{\cal O} }

\newcommand{\be}{\begin{equation}}
\newcommand{\ee}{\end{equation}}

\newcommand\rearth{{R$_{\oplus}$}}

\newcommand{\lhs}{LHS\,1140}
\newcommand{\lhsb}{LHS\,1140\,b}
\newcommand{\lhsc}{LHS\,1140\,c}
\newcommand{\lhsd}{LHS\,1140\,d}
\newcommand{\lfour}{L$_{\rm 4}$}
\newcommand{\lfive}{L$_{\rm 5}$}
\newcommand{\lflf}{\lfour{}/\lfive{}}

\usepackage{romannum}

\begin{document} 

%%%%%%%%%%%%%%%%%%%%%%%%%%%%%%%%%%%%%%%%%%%%

   \title{Planetary system LHS 1140 revisited \\ with ESPRESSO and TESS}

   \author{
%%--------------Lead Authors-----------------------
   J.~Lillo-Box\inst{\ref{cab}}, %\orcidicon{0000-0003-3742-1987}, }
%------------Contributing authors -----------------
   P.~Figueira\inst{\ref{eso},\ref{iace}},
   A.~Leleu\inst{\ref{cheops}},
   L.~Acu\~na\inst{\ref{marseille}},
   J.P.~Faria\inst{\ref{iace}}\fnmsep\inst{\ref{depfisporto}},
   N.~Hara\inst{\ref{geneva}},
   N.C.~Santos\inst{\ref{iace}}\fnmsep\inst{\ref{depfisporto}}, 
   A.~C.~M.~Correia\inst{\ref{coimbra}}\fnmsep\inst{\ref{obsparis}},%\\
   P.~Robutel\inst{\ref{obsparis}},
   M.~Deleuil\inst{\ref{marseille}},
   D.~Barrado\inst{\ref{cab}},
   S.~Sousa\inst{\ref{iace}},
%------------Others -----------------
   X.~Bonfils\inst{\ref{grenoble}},
   O.~Mousis\inst{\ref{marseille}},
   J.M.~Almenara\inst{\ref{grenoble}},%\\
   N.~Astudillo-Defru\inst{\ref{catolica}},
   E.~Marcq\inst{\ref{sorbonne}},
   S.~Udry\inst{\ref{geneva}},
   C.~Lovis\inst{\ref{geneva}},
   F.~Pepe\inst{\ref{geneva}}
}

\institute{
Centro de Astrobiolog\'ia (CAB, CSIC-INTA), Depto. de Astrof\'isica, ESAC campus 28692 Villanueva de la Ca\~nada (Madrid), Spain\label{cab} \email{Jorge.Lillo@cab.inta-csic.es  }
\and European Southern Observatory, Alonso de Cordova 3107, Vitacura, Region Metropolitana, Chile \label{eso}
\and Instituto de Astrof\' isica e Ci\^encias do Espa\c{c}o, Universidade do Porto, CAUP, Rua das Estrelas, PT4150-762 Porto, Portugal \label{iace} 
\and Physics Institute, Space Research and Planetary Sciences, Center for Space and Habitability - NCCR PlanetS, University of Bern, Bern, Switzerland \label{cheops}
\and Aix Marseille Univ, CNRS, CNES, LAM, Marseille, France \label{marseille} 
\and Depto. de F\'isica e Astronomia, Faculdade de Ci\^encias, Universidade do Porto, Rua do Campo Alegre, 4169-007 Porto, Portugal \label{depfisporto}
\and Geneva Observatory, University of Geneva, Chemin des Mailettes 51, 1290 Versoix, Switzerland \label{geneva}
\and CFisUC, Department of Physics, University of Coimbra, 3004-516 Coimbra, Portugal \label{coimbra}
\and IMCCE, Observatoire de Paris, PSL University, CNRS, Sorbonne Universite\'e, 77 avenue Denfert-Rochereau, 75014 Paris, France\label{obsparis}
\and CNRS, IPAG, Universit\'e Grenoble Alpes, 38000 Grenoble, France \label{grenoble}
\and Departamento de Matem\'atica y F\'isica Aplicadas, Universidad Cat\'olica de la Sant\'isima Concepci\'on, Alonso de Rivera 2850, Concepci\'on, Chile \label{catolica}
\and LATMOS/CNRS/Sorbonne Universit\'e/UVSQ, 11 boulevard d'Alembert, Guyancourt, F-78280, France \label{sorbonne}
}

   \date{Accepted for publication in A\&A on 24 September 2020}

% \abstract{}{}{}{}{} 
% 5 {} token are mandatory
 
  \abstract
   {\lhs{} is an M dwarf known to host two transiting planets at orbital periods of 3.77 and 24.7 days. They were detected with HARPS and Spitzer. The external planet (\lhsb{}) is a rocky super-Earth that is located in the middle of the habitable zone of this low-mass star. All these properties place this system at the forefront of the habitable exoplanet exploration, and it therefore constitutes a relevant case for further astrobiological studies, including atmospheric observations.}
   {We further characterize this system by improving the physical and orbital properties of the known planets, search for additional planetary-mass components in the system, and explore the possibility of co-orbitals.}
   {We collected 113 new high-precision radial velocity observations with ESPRESSO over a 1.5-year time span with an average photon-noise precision of 1.07~m/s. We performed an extensive analysis of the HARPS and ESPRESSO datasets and also analyzed them together with the new TESS photometry. We analyzed the Bayesian evidence of several models with different numbers of planets and orbital configurations.}
   {We significantly improve our knowledge of the properties of the known planets \lhsb{} ($P_b\sim24.7$~days) and \lhsc{} ($P_c\sim3.77$~days). We determine new masses with a precision of  6\%  for \lhsb{} ($6.48 \pm 0.46$~\Mearth{}) and 9\% for \lhsc{} ($m_c=1.78 \pm 0.17$~\Mearth{}). This reduces the uncertainties relative to previously published values by half. Although both planets have Earth-like bulk compositions, the internal structure analysis suggests that \lhsb{} might be iron-enriched and \lhsc{} might be a true Earth twin. In both cases, the water content is compatible to a maximum fraction of 10-12\% in mass, {which is equivalent to} a deep ocean layer of $779 \pm 650$~km for the habitable-zone planet \lhsb{}. Our results also provide evidence for a new planet candidate in the system ($m_d= 4.8\pm1.1$\Mearth{}) on a  ~78.9-day orbital period, which is detected through three independent methods. The analysis also allows us to discard other planets above 0.5~\Mearth{} for periods shorter than 10 days and above 2~\Mearth{} for periods up to one year. Finally, our co-orbital analysis discards co-orbital planets in the tadpole and horseshoe configurations of \lhsb{} down to 1~\Mearth{} with a 95\% confidence level (twice better than with the previous HARPS dataset). Indications for a possible co-orbital signal in \lhsc{} are detected in both radial velocity {(alternatively explained by a high eccentricity)} and photometric data {(alternatively explained by systematics), however}.}
   {The new precise measurements of the planet properties of the two transiting planets in \lhs{} as well as the detection of the planet candidate \lhsd{} make this system a key target for atmospheric studies of rocky worlds at different stellar irradiations.}


\titlerunning{LHS\,1140 revisited by ESPRESSO and TESS}
\authorrunning{Lillo-Box et al.}

   \maketitle
%

%===================================================
\section{Introduction}
%===================================================

In the past decades, the exploration of exoplanets has moved from the detection scheme to the characterization challenge. The new dedicated ground- and space-based facilities built for this purpose now offer the possibility of fully characterizing the properties of extrasolar planets with exquisite precision \citep[e.g., ][]{pepe14,suarez-mascareno20,damasso20}. {Understanding planet formation and evolution, determining atmospheric properties, and ultimately searching for biosignatures requires a sufficiently large sample of planetary systems. The plethora of properties of the current exoplanet population (more than 4100 known so far, according to the NASA Exoplanet Archive, \citealt{akeson13}) indeed offers a large collection of targets to carry out this endeavor}. The combination of different techniques then becomes critical to provide a complete view of the system, which in turn allows the inference of its history through feeding population synthesis models and allowing subsequent atmospheric characterization campaigns. In particular, the combination of the transits and radial velocity techniques is key to understanding the bulk composition of the planets. Exquisite photometric and radial velocity precision is needed to infer its internal structure, however, and key to setting observational constraints on formation and evolution processes.

Planets around low-mass stars have drawn the attention of the community in the recent years due to the ground-based instrumental capabilities in reaching the rocky domain \citep[e.g.,][]{luque19, zechmeister19}, also in the temperate region around the star \citep[TRAPPIST-1][]{gillon16}. The \lhs{} planetary system is one of these examples. This M4.5 dwarf is located 10.5 parsec away \citep{gaia18} and hosts two known planets, a small short-period telluric component (\lhsc{}, P$_{\rm c}\sim$3.77 days) and a temperate rocky super-Earth (\lhsb{}, P$_{\rm b}\sim$24.7 days). \lhsb{} was detected by \cite{dittmann17} using MEarth \citep{nutzman08} photometric time series and HARPS \citep{mayor03} radial velocities. 
Additional HARPS measurements and observations from the Spitzer Space Telescope by \cite{ment19} led to the detection of the inner component in the system, \lhsc{}, which also transits its host star. 

\lhsb{} lies within the habitable zone of the star and possesses a rocky composition, thus representing a key target for further astrobiological studies. {The path toward a habitability analysis involves, among others, the study of the internal structure of the planet \citep{shahar19}. Its composition and distribution is a direct consequence of its formation (e.g., the abundances of the different elemental building blocks) and evolution (i.e., the different heating and cooling processes, and impacts throughout the planet history), see \cite{dorn18} and references there in. The two planets in \lhs{} with their different orbital (hence irradiation) properties offer a unique opportunity to understand different evolutionary paths in the same environment.}

Added to this, \lhs{} is also an ideal target for searches of co-orbital planets. Co-orbital configurations consist of planet pairs trapped in gravitationally stable regions and 1:1 mean motion resonances. The Lagrangian points \lflf{} present in the gravitational field of a two-body system (like a star and a planet) are stable points of equilibrium that are located exactly on the same orbit as the planet, but $\pm 60^{\circ}$ ahead and behind it. These gravitational wells have been demonstrated to be very stable once an object is trapped \citep{laughlin02}. The only condition for stability, once trapped, is that the total mass of the planet pair (planet plus trojan) must be lower than 3.8\% of the mass of the star \citep[e.g.,][]{gascheau1843}. This relaxed constraint allows similar-mass planets to co-orbit together in a long-term stable dance around the star. The two planets can share the same orbital path in different co-orbital configurations. In the case of circular or quasi-circular orbits, the possibilities are \textit{\textup{tadpole}} and \textit{\textup{horseshoe}} orbits. In the first case, the co-orbital surrounds (or librates around) one of the Lagrangian points in the co-rotating frame with the planet, as in the case of the Jupiter trojans. In the horseshoe configuration, the co-orbital describes a horseshoe shape in the co-rotating frame, moving from \lfour{} to \lfive{} and exchanging orbits with the planet. This is the case, for instance, of the Saturnian moons Janus and Epimetheus, which have comparable sizes. In the case of eccentric orbits, other configurations come into play, such as the quasi-satellite or the anti-Lagrange configuration \citep{giuppone10}. In general, several studies have shown that the formation and stability of co-orbital planets makes these configurations not only possible but probable \citep{cresswell08,cresswell09,leleu19} and that their dynamical properties could have hidden them in the noise of already existing data \citep{ford07,madhusudhan08,janson13,hippke15}. However, despite the ample room for stability estimated from theoretical studies, no co-orbital pairs have been found so far. Several attempts have been focused on these configurations and have so far set observational constraints to their existence in different regimes \citep[see, e.g., ][]{lillo-box18a, lillo-box18b}{, and} one key candidate has already been found (TOI-178, \citealt{leleu19}). 

The \lhs{} planetary system has important properties that make it a key system in which to search for these co-orbitals. First, the low mass of the host star allows reaching the sub-Earth mass domain. The transiting nature of the two known planets allows the application of co-orbital detection techniques, thus avoiding the degeneracy with the eccentricity at first order \citep{leleu17}. The edge-on orientation of the planetary system also permits the search for co-orbitals through the transit technique \citep{janson13,lillo-box18b}. The slow rotational velocity of the star allows precise radial velocity measurements. {Moreover,} the multi-planet nature of the system increases the likelihood  that it hosts co-orbital pairs \citep{leleu19b}, especially when the planets are in mean motion resonances \citep{cresswell08,leleu19b}. 

We present the joint analysis of the first observations of this system obtained with the Echelle SPectrograph for Rocky Exoplanets and Stable Spectroscopic Observations (ESPRESSO) instrument and the precise light curve obtained with the Transiting Exoplanets Survey Satellite (TESS) mission. The data are presented in Sect.~\ref{sec:observations}. In Sect.\ref{sec:rv_explore} we explore the new radial velocities in search for other planets in the system. The TESS light curve is also further explored in Sect.~\ref{sec:lc_explore}. In Sect.~\ref{sec:coorbitals} we explore the possibility of co-orbitals to the two known planets in the system from different perspectives and with different techniques. The final joint data analysis including the radial velocity and the light curve is presented in Sect.~\ref{sec:joint}. We discuss the result in Sect.~\ref{sec:discussion} and provide the final conclusions in Sect.~\ref{sec:conclusions}.

%===================================================
\section{Observations}
%===================================================
\label{sec:observations}

%--------------------------------------------------------------------
\subsection{HARPS}
\label{sec:harpsRVs}

\lhs{} was intensively observed by  with the High Accuracy Radial velocity Planet Searcher (HARPS, \citealt{mayor03}) at the 3.6\,m telescope of the European Southern Observatory (ESO) La Silla facilities under the program IDs  191.C-0873 and 198.C-0838 (PI: X. Bonfils) and 0100.C-0884 (PI: N. Astudillo-Defru). In total, this radial velocity dataset comprised 293 HARPS radial velocity measurements that were presented in \cite{ment19}, spanning 783 days between  November 2015 ($JD=2457349.65$) and January 2018 ($JD=2458133.54$). The average cadence was one epoch every 1.8 days, but normally, two spectra were obtained every night with separations of a few hours. We performed a night binning of the dataset, as suggested in \cite{dumusque12}. This  simple strategy has proved to be efficient to reduce the effect of short-term correlated noises on orbital elements \citep[e.g., ][]{hara19}. In total, 145 individual measurements were available. The corresponding uncertainties per binned data point are distributed around $3.2\pm1.2$~m/s. The periodogram of this dataset is presented in Fig.~\ref{fig:periodograms}.

%--------------------------------------------------------------------
\subsection{ESPRESSO}
\label{sec:espressoRVs}

ESPRESSO is the new ultra-stable high-resolution spectrograph of the Very Large Telescope at ESO's Paranal Observatory \citep{pepe20}. This facility has the capability of collecting light from any of the four 8.2 m Unit Telescopes (UTs). The instrument has two arms with a total wavelength coverage of 380 to 788 nm. ESPRESSO is equipped with a powerful reduction pipeline that performs all steps of the basic reduction and provides high-level data products, including radial velocities. The aim is a final radial velocity precision down to the 10 cm/s level.

\begin{figure*}
\centering
\includegraphics[width=\textwidth]{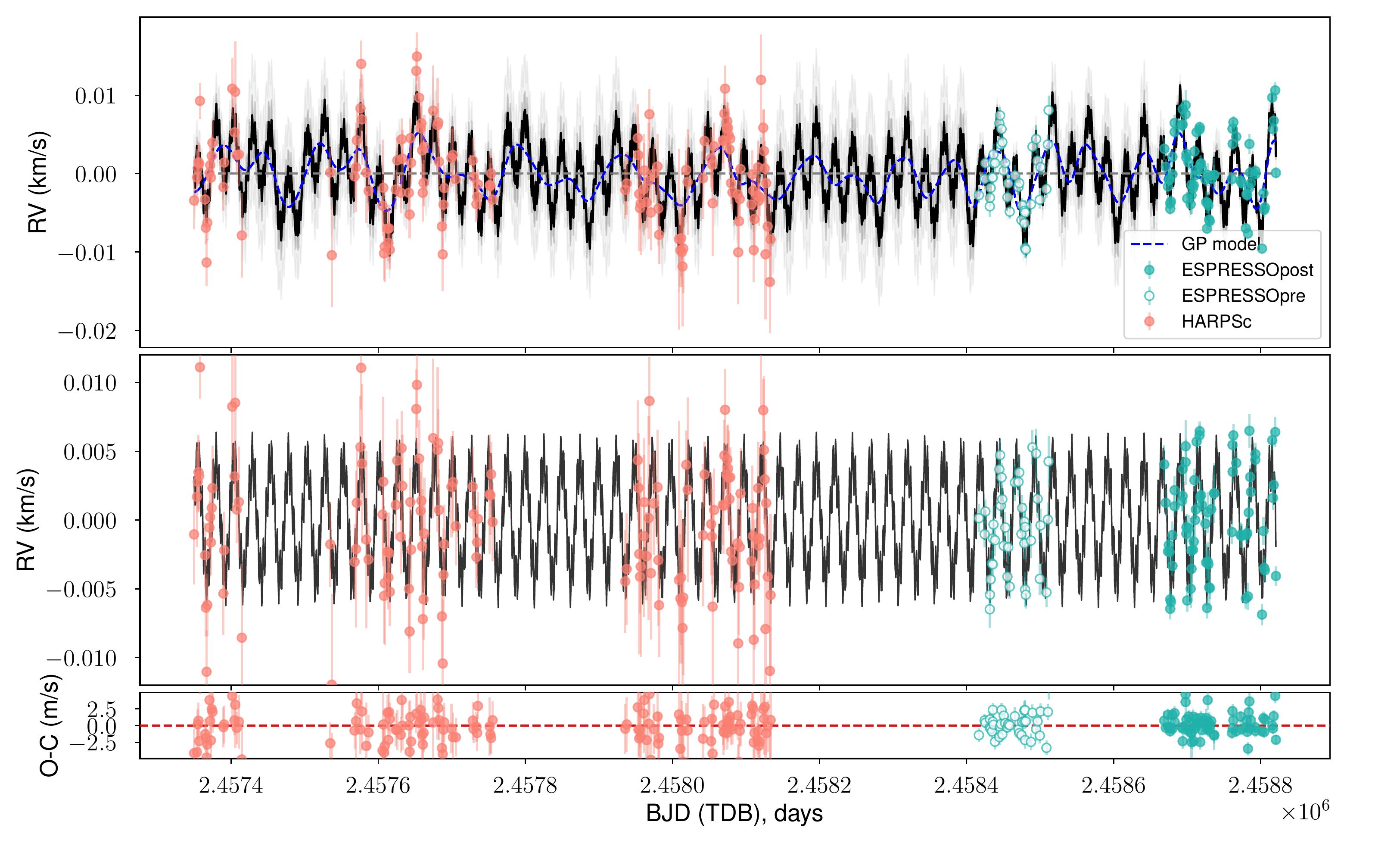}
\caption{\textbf{Top:} Radial velocity of \lhs{} from the HARPS (red) and ESPRESSO (open for  ESPRESSOpre and filled for ESPRESSOpost in green) datasets. The black line shows the median radial velocity model from the joint photometric and radial velocity analysis including the two known Keplerian signals and the GP model (see Sect.~\ref{sec:joint}). The gray shaded regions correspond to the 68.7\% (dark gray) and 95\% (light gray) confidence intervals of the model. The median GP model is shown as a dashed blue line. \textbf{Middle:} Radial velocity dataset after removing the median GP model. The Keplerian model is shown as a solid black line. \textbf{Bottom:} Radial velocity residuals of the full model.}
\label{fig:rvtime}
\end{figure*}

\begin{figure*}
\centering
\includegraphics[width=0.48\textwidth]{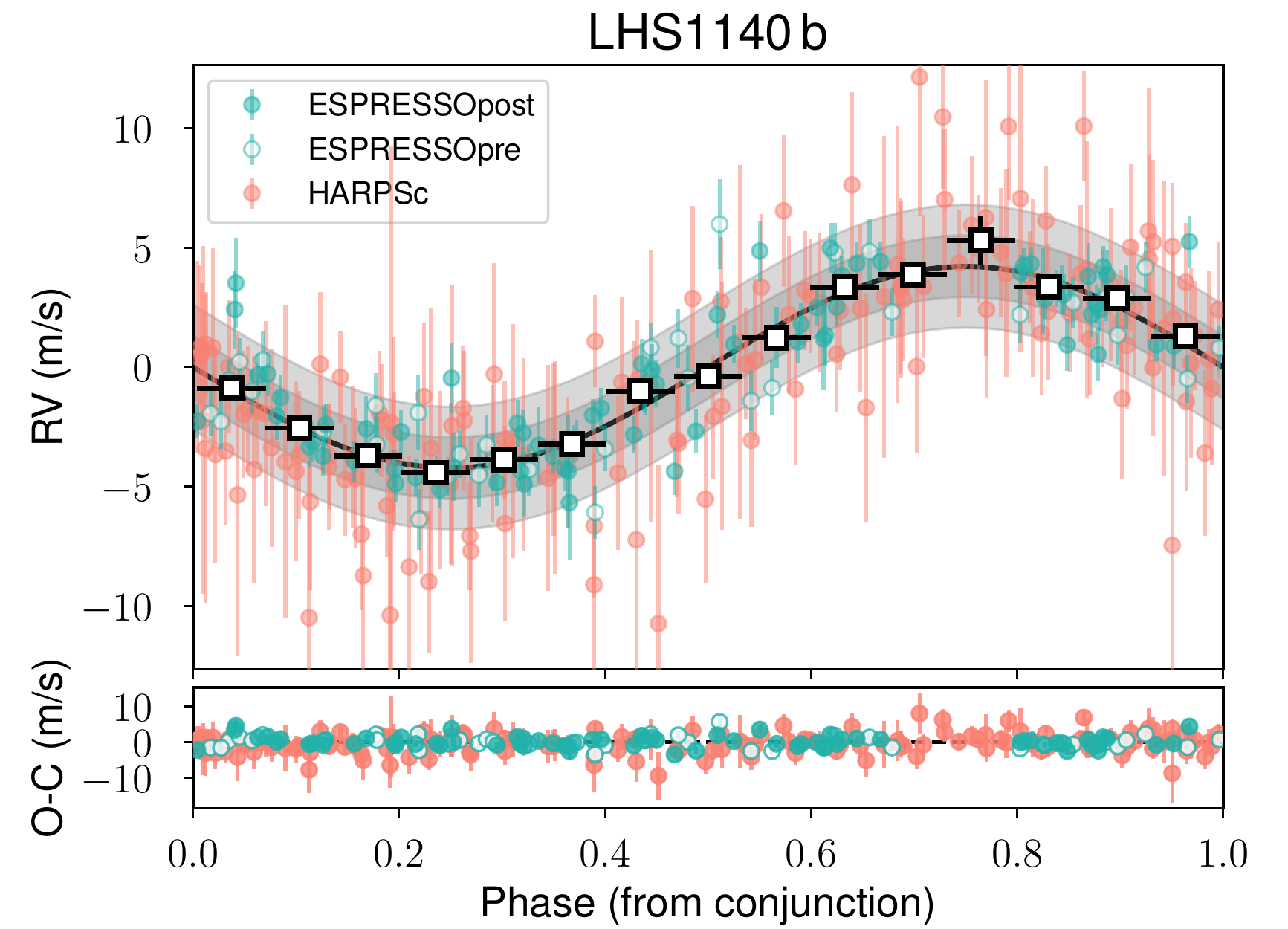}
\includegraphics[width=0.48\textwidth]{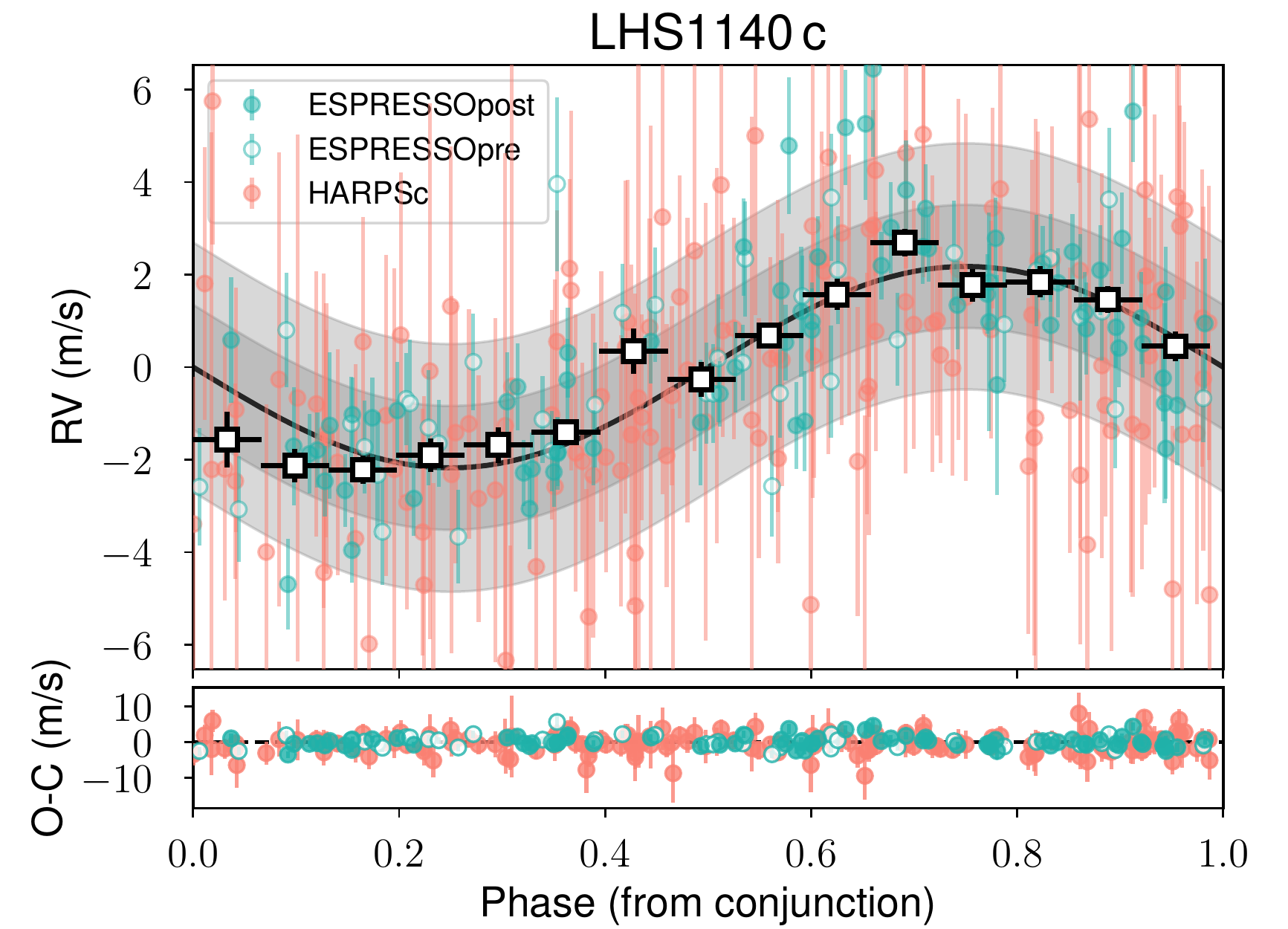}
\caption{Phase-folded radial velocity signal of \lhsb{} (left panels) and \lhsc{} (right panels). For each planet, the upper panel shows the radial velocity data (color-coded as in the legend), binned radial velocities (open black symbols) with a bin size corresponding to 15\% of the phase, the median model for each planet obtained from the joint fit analysis (see Sect.~\ref{sec:joint}) as the solid black line, and the 68.7\% and 95\% confidence intervals as shaded dark and light gray regions. The bottom panels show the residuals of the median model.}
\label{fig:rvphase}
\end{figure*}

We obtained 116 spectra of \lhs{} in three consecutive semesters\footnote{Prog. IDs: 0102.C-0294(A) , 0103.C-0219(A)  and 0104.C-0316(A), PI: J. Lillo-Box} (P103, P104, and P105). The total time span of the observations is 404 days; they were taken between October 2018 ($JD=2458416.71$) and December 2019 ($JD=2458820.56$), and a mean cadence of one spectrum was obtained every 2.2 days (median of one spectrum per night). Each spectrum was obtained with an exposure time of 1820\,s in P102 and 1915\,s in P103 and P104, and produced a median signal-to-noise ratio (S/N) $\rm{S/N}=80$ at 700\,nm. We processed the whole dataset using version v2.0.0 of the ESPRESSO Data reduction Software\footnote{\url{https://www.eso.org/sci/software/pipelines/espresso/espresso-pipe-recipes.html}} (DRS) pipeline \citep{pepe20}. We selected a binary mask corresponding to an M5 spectral type for this M4.5 star to perform the cross-correlation and obtain the final radial velocities \citep{baranne96,pepe03}. The resulting radial velocities have a mean photon-noise uncertainty of 1.07 m/s. In June 2019, ESPRESSO suffered a major intervention, and the chamber and the vacuum vessel were opened to perform a fiber-link exchange. This introduced a small jump in the radial velocity datasets taken before and after the intervention, and  the actual value of this jump is not constant for the different spectral types. The data taken before (labeled ESPRESSOpre in this paper) and after (ESPRESSOpost) should therefore be treated as coming from different instruments. This fiber-link exchange occurred between our first and second observation semester. The mean precision corresponding to the data obtained before the fiber-link exchange is 1.16 m/s, and the post-change data have an average uncertainty of 1.02 m/s. This improvement goes in line with the increase in transmission measured after the intervention \citep{pepe20} and a slight increase in exposure time from 1820\,s in P102 to 1915\,s in P103 and P104. We removed two data points with uncertainties above 3 m/s because low clouds caused a low S/N (the observations were aborted after half the requested exposure time). The final dataset including the radial velocities, activity indicators, S/N, and exposure time for each spectrum is presented in Table~\ref{tab:espresso}. Figure~\ref{fig:rvtime} shows the time-series radial velocity for both HARPS and ESPRESSO, and Fig.~\ref{fig:rvphase} shows the phase-folded curves for the two known planets in the system.

\begin{figure}
\centering
\includegraphics[width=0.5\textwidth]{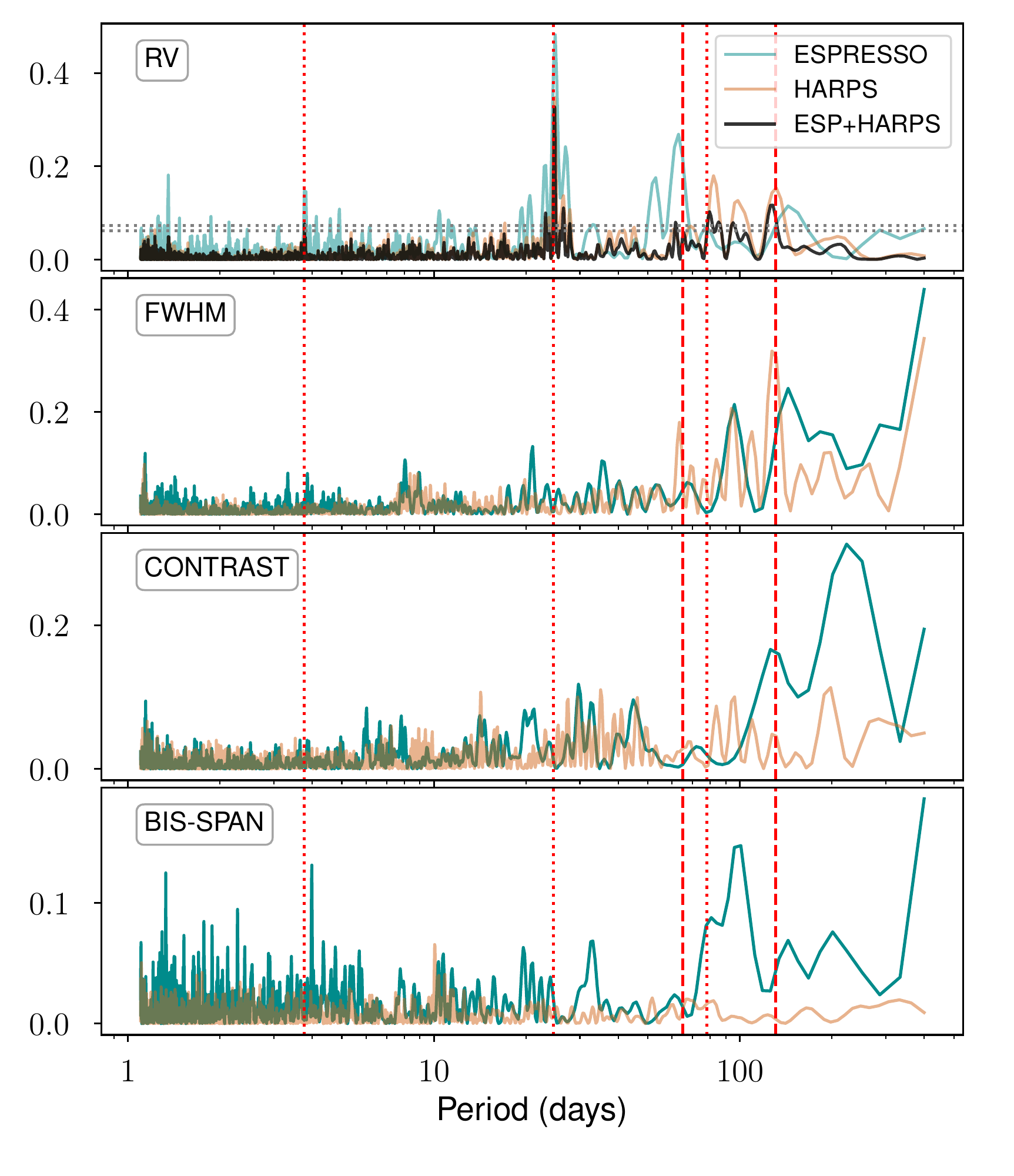}
\caption{Periodogram of the radial velocity (upper panel) for each dataset individually (ESPRESSO in light green and HARPS in red) and for the joint dataset of both instruments assuming a 26 m/s offset (see Sect.~\ref{sec:joint}). The activity indicators for the individual datasets are also shown in the lower panels. The periods of the three planets are marked as dotted vertical lines, and the rotation period (P$_{\rm rot}=131$~days) and its first harmonic (P$_{\rm rot}/2=65$~days) are marked as vertical dashed lines.}
\label{fig:periodograms}
\end{figure}

The periodogram of the full dataset (including HARPS and ESPRESSO measurements, see Fig.~\ref{fig:periodograms}) shows clear signals of the known planets \lhsb{} and \lhsc{}. The two signals are detectable independently in the two instrument datasets (ESPRESSO and HARPS), and their significance is boosted when they are combined by only including a 26 m/s radial velocity offset between the two (see Sect.~\ref{sec:rv3p}). In addition to the signals of these two planets, other signals appear at a significant level. The rotation period of the star at $P_{\rm rot}=131$ days is clearly detected in each of the two datasets and in the combined set. Interestingly, both datasets also show a peak at half the rotation period $P_{\rm rot}/2\sim65$~days, which is not as evident in the HARPS radial velocity dataset. However, the rotation period and its alias are clearly visible in the activity index corresponding to the full width at half maximum (FWHM) of the cross-correlation function (CCF) in both datasets, see Fig.~\ref{fig:periodograms}. Added to this, additional signals stand out in the region around 70-100 days. This is further investigated in Sect.~\ref{sec:rv3p}. 

Taking advantage of the large number of ESPRESSO spectra, we used ODUSSEAS \citep{antoniadis20} to estimate the effective temperature and metallicity of this M star. This machine-learning tool measures pseudo-equivalent widths for more than 4000 spectral lines and compares them to a training dataset composed of HARPS spectra of reference M-dwarf stars. Although the code was originally tested for spectra with resolutions from 48\,000 to 115\,000, we used the combined ESPRESSO spectrum directly with a higher resolution of 140\,000. Although the resolution is slightly higher than the highest resolution grid of the code (115\,000), we can safely assume that the method that measures the pseudo-equivalent widths provides compatible values when a spectrum with higher resolution is used\footnote{{This has been tested with observations from the ESPRESSO Guarantee Time Observations (PI: F. Pepe).}}. With this code we derived an effective temperature of $T_{\rm eff}=2988\pm67$~K and a metallicity of ${\rm [Fe/H]}=-0.262\pm0.104$~dex. These errors were estimated by considering on the one hand the machine-learning model error (precision errors of 17~K and 0.03~dex, respectively), and on the other hand the mean absolute errors of the machine-learning models for the training dataset (65~K and 0.10~dex for a resolution of 115\,000 for the dataset grid,  \citealt{antoniadis20}).

%--------------------------------------------------------------------
\subsection{TESS photometry}

\begin{figure}
\centering
\includegraphics[width=0.5\textwidth]{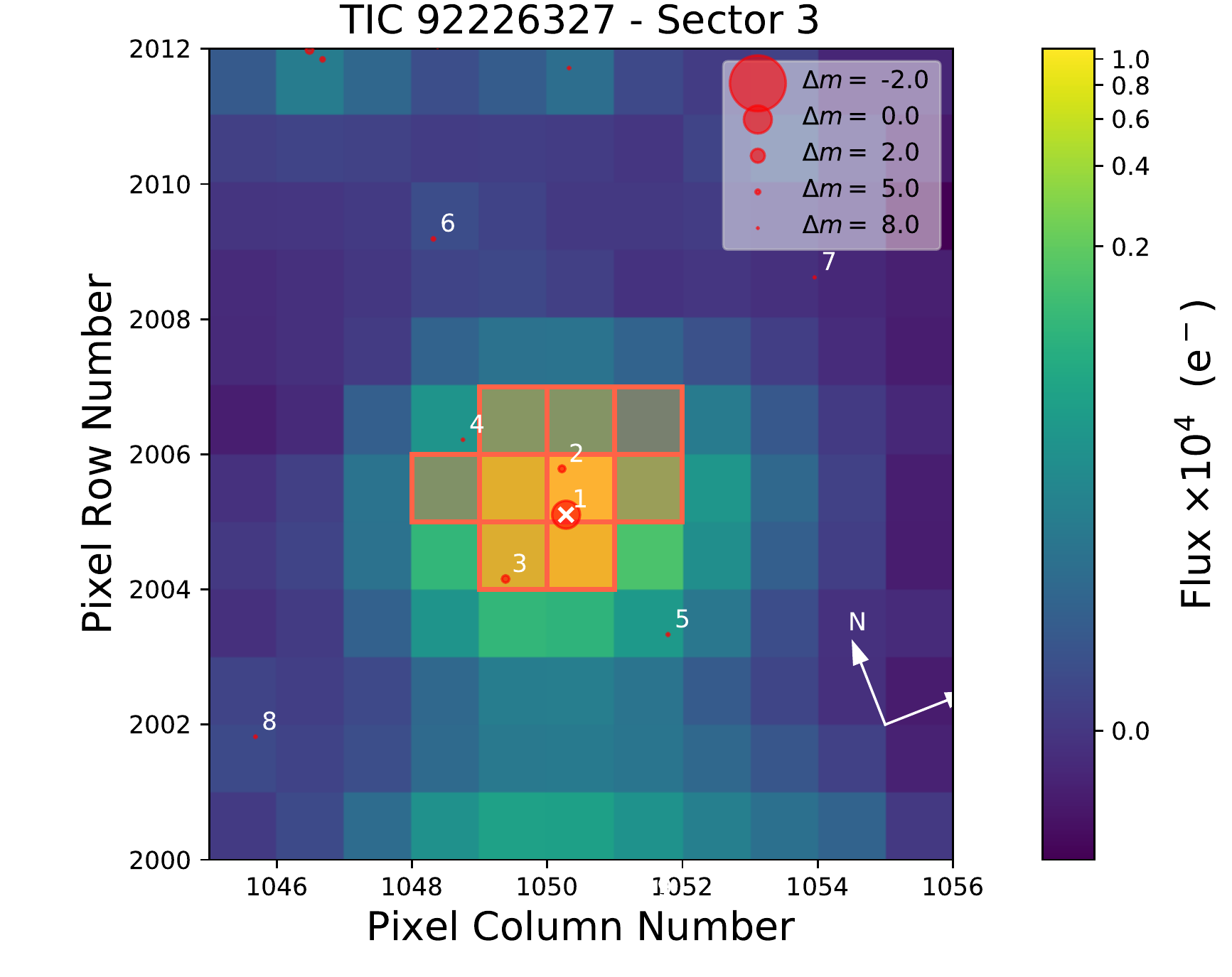}
\caption{The TPF of \lhs{} from the TESS observations in Sector 3 (composed with \texttt{tpfplotter}, \citealt{aller20}). The SPOC pipeline aperture is overplotted with shaded red squares, and the Gaia DR2 catalog is also overlaid with symbol sizes proportional to the magnitude contrast with the target, marked with a white cross.}
\label{fig:tpf}
\end{figure}

The Transiting Exoplanet Sky Survey (TESS, \citealt{ricker14}) observed \lhs{} during Sector 3 in camera 1, from 20 September 2018 ($JD=2459115$) to 18 October 2018 ($JD=2459141$), immediately before our ESPRESSO campaign. We used \texttt{tpfplotter}\footnote{\url{https://github.com/jlillo/tpfplotter}} \citep{aller20}to check for contaminant sources in the automatically selected aperture. This is shown in Fig.~\ref{fig:tpf}, where we display all sources from the Gaia DR2 catalog with magnitude contrast up to $\Delta m=8$~mag.  In addition to \lhs{}, two additional sources lie inside the TESS aperture with magnitude contrasts in the Gaia passband of 3.8 and 4.0 mag. This imposes an upper limit to the dilution factor of 5.2\%.

\begin{figure*}
\centering
\includegraphics[width=1\textwidth]{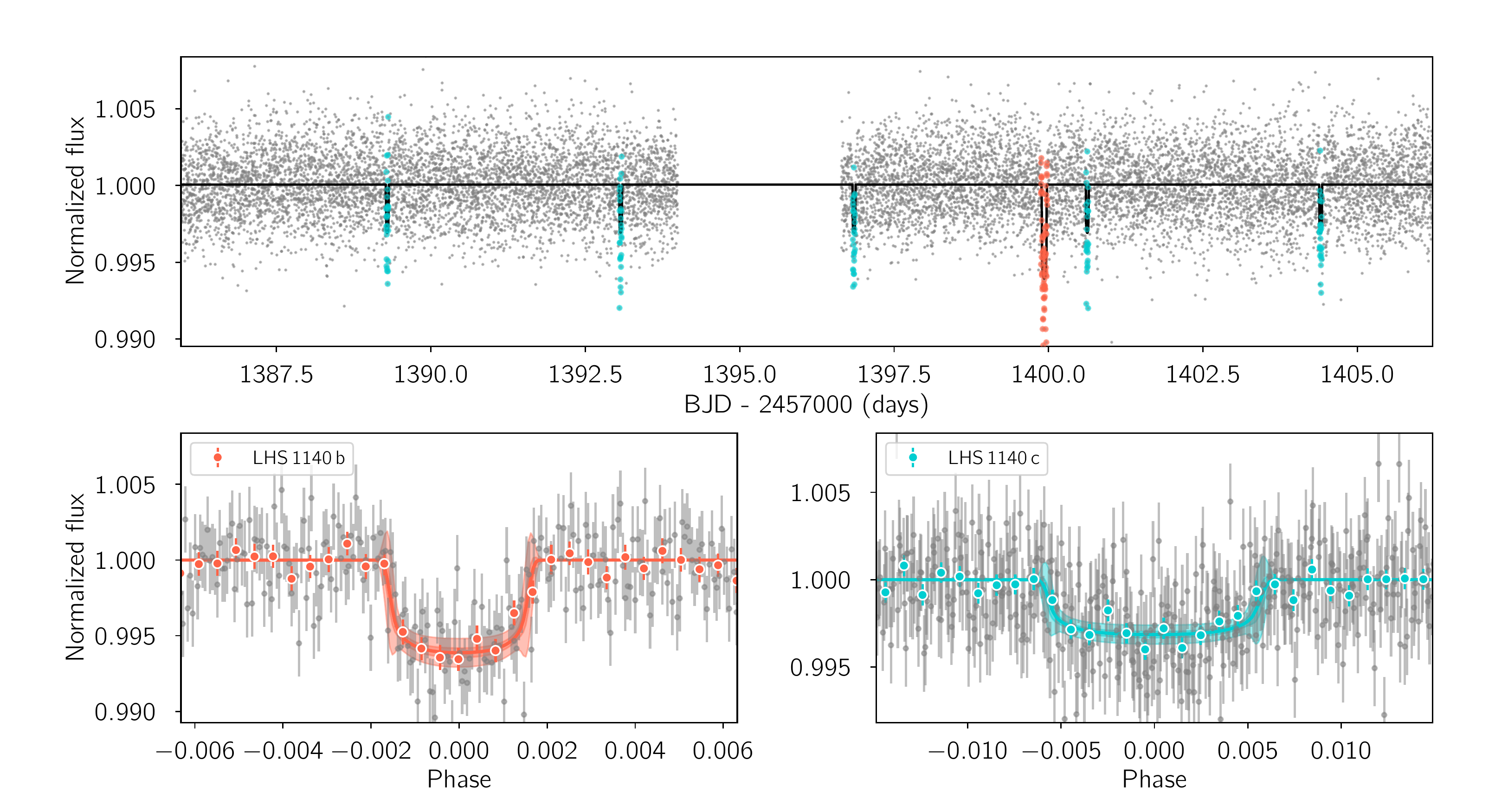}
\caption{TESS light curve extracted from the SPOC pipeline and detrended using the PDCSAP flux. \textbf{Upper panel:} Complete light curve displaying the transits of the two known transiting planets in the system, \lhsb{} (red, one transit) and \lhsc{} (light green, five transits). \textbf{Lower panel:} Phase-folded light curve centered on the planet phase for each of the two transiting planets,  \lhsb{} (bottom left) and \lhsc{} (bottom right). The colored symbols correspond to bins of one fifth of the transit duration (i.e., 30 min for \lhsb{} and 11 min for \lhs{c}). }
\label{fig:tess}
\end{figure*}

We used the light curve extracted by the SPOC pipeline. For the purpose of this paper, we use the presearch data-conditioning simple aperture photometry (PSDCSAP) detrending of the data, which has a CDPP of 0.547 parts per thousand (hereafter ppt). In Fig.~\ref{fig:tess} we show the extracted photometric time series and the phase-folded light curves for the two known planets. {The photometric precision of this time series makes it sufficient for an independent analysis and characterization of the two planets. We therefore did not use the previous observations from Spitzer and MEarth presented in \cite{ment19}.}

%===================================================
\section{Exploring the radial velocity dataset}
%===================================================
\label{sec:rv_explore}

%-------------------------------------------------------------------------
\subsection{$\ell_1$ periodogram}
\label{sec:l1per}

We first analyzed the radial velocity data with the $\ell_1$-periodogram, as defined in~\cite{hara17}. This  tool is designed to search for a representation of the signal as a sum of a small number of sinusoids, where ``small'' is compared to the number of observations. It has a similar aspect to a regular periodogram, but with far fewer peaks due to aliasing. As in~\cite{hara20}, we computed the $\ell_1$-periodogram of the data with different assumptions on the noise covariance. The covariance models were then ranked by cross-validation. We {considered} the ESPRESSO and HARPS data without binning. 

\begin{figure}
\centering
\includegraphics[width=0.5\textwidth]{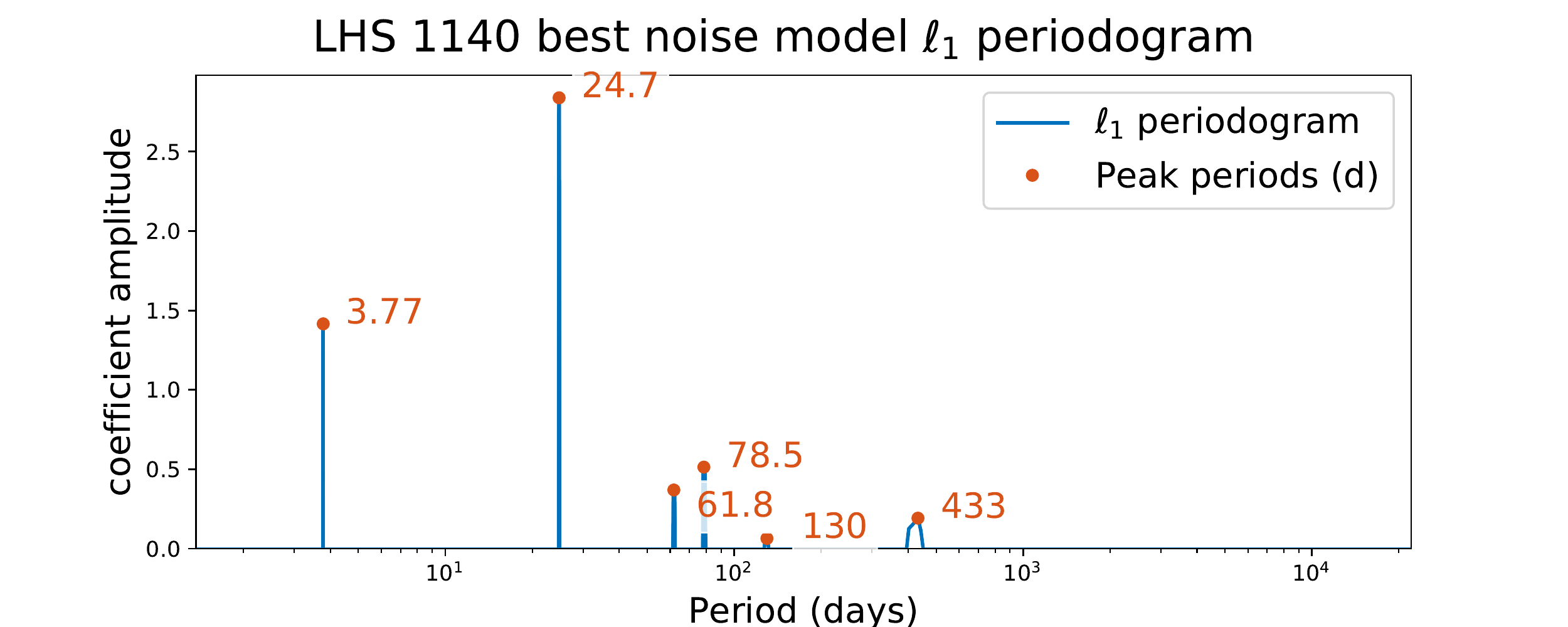}
\caption{$\ell_1$-periodogram of the LHS 1140 HARPS and ESPRESSO data with the noise model that has the best cross-validation score.}
\label{fig:l1p}
\end{figure}

Similarly to~\cite{haywood14}, we included in the model two activity indicators per instrument, smoothed with Gaussian kernels with different timescales, as linear predictors. These indicators were chosen because they exhibited  significant variations on a short (0-2 days) and longer timescale (order of days and 10 - 20 days).  We included the FWHM, with a smoothing timescale of 20 days and 2 days for HARPS and ESPRESSO, respectively, the pipeline-derived line asymmetry for ESPRESSO (10 days) and bisector span for HARPS (1 day). The timescales were chosen after fitting the hyperparameters of the Gaussian kernel. We further added a quadratic trend to the model. We assumed a noise model with a white-noise component $\sigma_W$, a calibration error $\sigma_C$ , and a Gaussian component with amplitude $\sigma_R$ and exponential decay $\tau$. We then considered all possible combinations of values for 
$\sigma_R$ = 0.0 ,0.5, 1.0, 1.5, 2 m/s, 
$\sigma_W$ = 0.5, 1 m/s
$\sigma_C$ = 0.1, 0.5, 0.75 m/s, and
$\tau$ = 0.0, 1.0, 3.0, 6.0 d. 
All these noise models were ranked with cross validation, as in~\cite{hara20}. In Fig.~\ref{fig:l1p} we present the $\ell_1$ periodogram obtained with the noise model with maximum cross-validation score (corresponding to $\sigma_W$ = 1 m/s,  $\sigma_R$ = 0.5 m/s, $\tau$ = 3 d, and   $\sigma_C$ = 0.75 m/s.).

We then considered the 20\% highest ranked model $\mathrm{CV}_{20}$, and computed the number of times that a signal is included in the model (which reaches a false-alarm probability, ${\rm FAP}<0.05$). We find that signals might be included in the models at seven periods. In decreasing order of significance, these are at 24.7, 3.77, 78.8, 62, 424, 129, and  16.9 days (see Table~\ref{tab:cv}). The last two signals are included only in 50\% of the $\mathrm{CV}_{20}$ models, and  16.9 days does not appear in the highest ranked noise model. The other signals appear to be present in the data. The signals at 3.77 and 24.7 days correspond to the known planets \lhsc{} and \lhsb{}, respectively. The signals at 130 days and 62 days are likely linked to the rotation period of the star (P$_{\rm rot}$) and half of it, P$_{\rm rot}/2$. The origin of the 433 days signal is less certain because its significance also depends on the way the stellar indicators are included in the model. The signal at 78.8 days (the third most significant in the $\ell 1$-periodogram) might be due to a planet and hence deserves additional attention. 

\begin{table}   
\centering
\begin{tabular}{p{1cm}|p{1.5cm}|p{1.5cm}|p{1.5cm}} 
Period (d) & FAP (best fit) & Inclusion in the model & $\mathrm{CV}_{20}$ median FAP \\ \hline \hline 
3.777& $1.17\cdot10^{-22}$ & 100.0\% & $1.37\cdot10^{-17}$ \\ 
16.94& $-$ & 45.0\% & \\  
24.75& $2.30\cdot10^{-30}$ & 100.0\% & $6.77\cdot10^{-39}$\\ 
61.97& $1.04\cdot10^{-4}$ & 100.0\% & $8.82\cdot10^{-11}$\\ 
78.79& $9.38\cdot10^{-8}$ & 100.0\% & $4.57\cdot10^{-8}$ \\ 
129.0& $4.19\cdot10^{-2}$ & 50.0\% & \\ 
424.0& $2.03\cdot10^{-4}$ & 100.0\% & $2.29\cdot10^{-5}$\\
\end{tabular} 
\caption{Signals appearing in the $\ell_1$ periodogram. From left to right, the columns are the signal periods, their FAP with the 20\% highest ranked model ($\mathrm{CV}_{20}$ models), the percentage of $CV_{20}$ models in which these signals are included (i.e., reach an FAP<0.05), and the median FAP in the $CV_{20}$ models.}
\label{tab:cv}
\end{table}

%-------------------------------------------------------------------------
\subsection{Evidence for a planet with a period of $\sim$80 days}
\label{sec:rv3p}

As reported in \cite{ment19} and confirmed in the new ESPRESSO dataset (see Sect.~\ref{sec:espressoRVs} and Sect.~\ref{sec:l1per}), the periodogram of the joint dataset shows a power excess between the first and second harmonic of the rotational period of the star; this is between 65 and 131 days. These several signals have a maximum peak at around 80 days. We here explore the evidence for this third signal by studying different scenarios (different number of planets, from one planet to three planets; and different orbital configurations, eccentric or circular) and comparing the models using Bayesian analysis to unveil the significance of this signal.

\begin{figure*}
\centering
\includegraphics[width=1\textwidth]{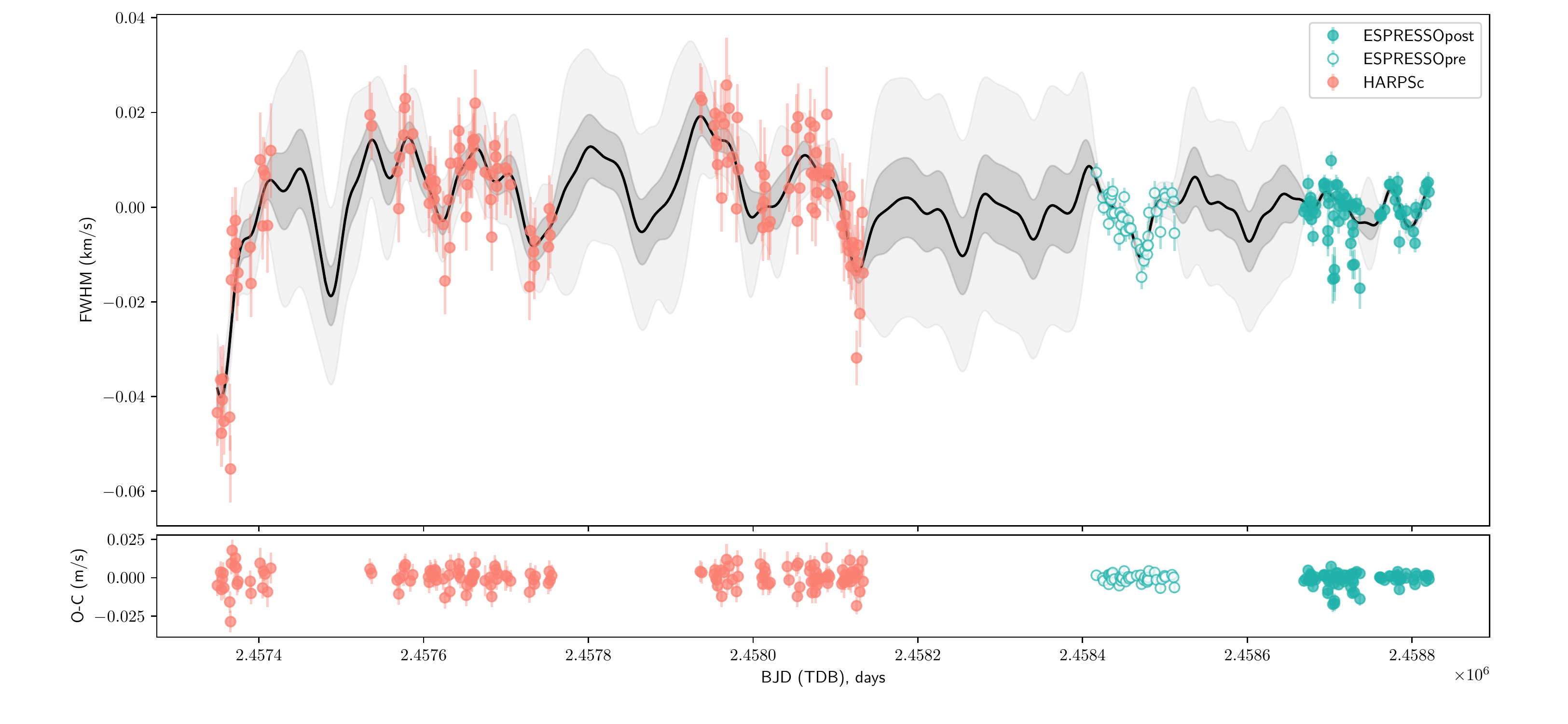}
\caption{Dataset of the FWHM of the cross-correlation function from the HARPS and ESPRESSO data. The systemic offset for each instrument as retrieved from the two-planet model has been removed. The median GP model is shown as the solid black line, and the 1$\sigma$ and 3$\sigma$ confidence intervals are shown as shaded dark and light gray regions, respectively. The bottom panel shows the residuals after removing the GP model.}
\label{fig:fwhm}
\end{figure*}

We explored each dataset separately (i.e., HARPS and ESPRESSO), together with a final full radial velocity dataset analysis (i.e., HARPS+ESPRESSO). Each planet signal was modeled as a Keplerian function, with independent values for the radial velocity semiamplitude of each planet ($K_i$), orbital period ($P_i$), eccentricity ($e_i$), and argument of the periastron ($\omega_i$). Because the clear effect of the stellar activity on the radial velocity is shown in the periodogram and has been pointed out by previous studies, we used Gaussian processes (GPs) to model the correlated noise. 
We used the quasi-periodic kernel from the \texttt{george} \citep{george} implementation, which is a combination of an exponential decay and a periodic part,~
\begin{eqnarray}
\Sigma_{ij} = \eta_1^2 \cdotp \mathrm{exp}\left[  - ~ \frac{(t_i-t_j)^2}{2\eta_2^2} ~ - ~ \frac{\sin^2{\frac{\pi (t_i-t_j)}{\eta_3}}}{\eta_4^2}   \right]
.\end{eqnarray}
~
Here $\eta_1$ corresponds to the amplitude of the correlated noise, $\eta_2$ can be interpreted as the timescale of the variations of the stellar features causing the correlated noise \citep[see, e.g.,][]{faria16}, $\eta_3$ represents the stellar rotation period (P$_{\rm rot}$), and $\eta_4$ is a balance between the exponential and the periodic components of the kernel. As shown in \cite{suarez-mascareno20}, we can additionally use the FWHM of the cross-correlation function as an activity indicator to further constrain the GPs (see the FWHM time series in Fig.~\ref{fig:fwhm}). In this case, the radial velocities and the FWHM share all hyperparameters of the GP, except for the amplitude, therefore an additional parameter per instrument ($\eta_{\rm 1,FWHM,j}$) is included. For ESPRESSO, which has the highest resolution of the planet searchers and the most stable instrumental profile,  the FWHM is expected to be the best tracker of line-deforming stellar activity. This indicator therefore provides an excellent way to determine the radial velocity variations that are caused by this stellar activity. For every instrument, we also added a jitter value for the radial velocity and the FWHM ($\sigma_{\rm RV,j}$ and $\sigma_{\rm FWHM,j}$, respectively) to account for additional uncorrelated noise and a systemic offset ($\Delta_{\rm RV,j}$ and $\Delta_{\rm FWHM,j}$). 

We used the \texttt{emcee}\footnote{\url{https://emcee.readthedocs.io/en/stable/}} \citep{foreman-mackey13} implementation of the \cite{Goodman2010} affine invariant Markov chain Monte Carlo (MCMC) ensemble sampler to explore the parameter space and sample the posterior distribution of each parameter. For each model, we used a number of walkers equal to four times the number of parameters involved (N$_p$) and 20\,000 steps for each walker. Following the recommendations from the \texttt{george} package, we first explored the parameter space with a full run using all the steps and walkers. In a second phase, we resampled the walker positions around the best walker and ran a second iteration with the same number of steps to improve convergence. A final run of the maximum probability region was produced by using 10\,000 steps per walker. Convergence was checked by estimating the autocorrelation time ($\tau$) of the chains and ensuring a minimum length of the chain, corresponding to $30\times\tau$. The final joint chain was then composed of $40\,000\times N_p$, which for the simplest model (no planets) corresponds to $6.8\times10^5$ steps and for the most complex model (including three Keplerians and three instruments) corresponds to $1.3\times10^6$ steps. 

The priors for each parameter involved in the different models are shown in Table~\ref{tab:rv3p}. In brief, we assumed Gaussian priors on the planet periods and mid-transit times, centered on the values from \cite{ment19}, but with a broad width of five times the published uncertainties. The remaining Keplerian parameters were set to broad uniform priors throughout the entire allowed regime. The prior on the GP hyperparameter $\eta_3$ was set to a Gaussian centered on the rotational period of the star, corresponding to $131\pm5$ days \citep{dittmann17}. Based on the observed radial velocities, we set the $\eta_1$ and $\eta_{\rm 1,FWHM}$ parameters to a uniform distribution with a maximum amplitude of 100 m/s. The timescale hyperparameter $\eta_2$ was set to a broad range of values between one and five times the stellar rotation period. The instrument systemic velocities were set up with uniform priors between -13.4 and -13.0 km/s, and the radial velocity jitter for each instrument was allowed to have values up to 20 m/s (based on our first model attempts).

In total, 15 models (including zero to three planets and all combinations of circular and eccentric orbits for each of them) were tested for each of the three datasets (HARPS, ESPRESSO, and HARPS+ESPRESSO), that is, 45 models were tested in total. In order to statistically compare the different models and datasets, we estimated the Bayesian evidence ($\mathcal{Z}$) of each model by using the \texttt{perrakis}\footnote{\url{https://github.com/exord/bayev}. This code is a \textit{python} implementation by R. D\'iaz of the formalism explained in \cite{perrakis14}.} implementation. Based on this Bayesian evidence, we can estimate the Bayes factor ($\mathcal{B}$) between two models as the ratio between the evidence of each pair of models as $\mathcal{B}=\ln{\mathcal{Z}_i}-\ln{\mathcal{Z}_j}$. The model with the strongest evidence has the highest statistical relevance, with $\mathcal{B}>6$ considered as a strong evidence for one model against the other. 

The three panels in Fig.~\ref{fig:evidence} display this Bayesian evidence for all 15 models considered in each dataset. We find that the two-planet model with both planets on circular orbits has the strongest evidence in all datasets. This means that the current data are unable to provide evidence for eccentric architectures of the two known planets. We can only place upper limits of $e_b<0.096$ and $e_c<0.274$ (at 95\% confidence level) based on the noncircular models. The median and 68.7\% confidence interval of the parameters for this two-planet model are presented in Table.~\ref{tab:rv3p}.

\begin{figure}
\centering
\includegraphics[width=0.5\textwidth]{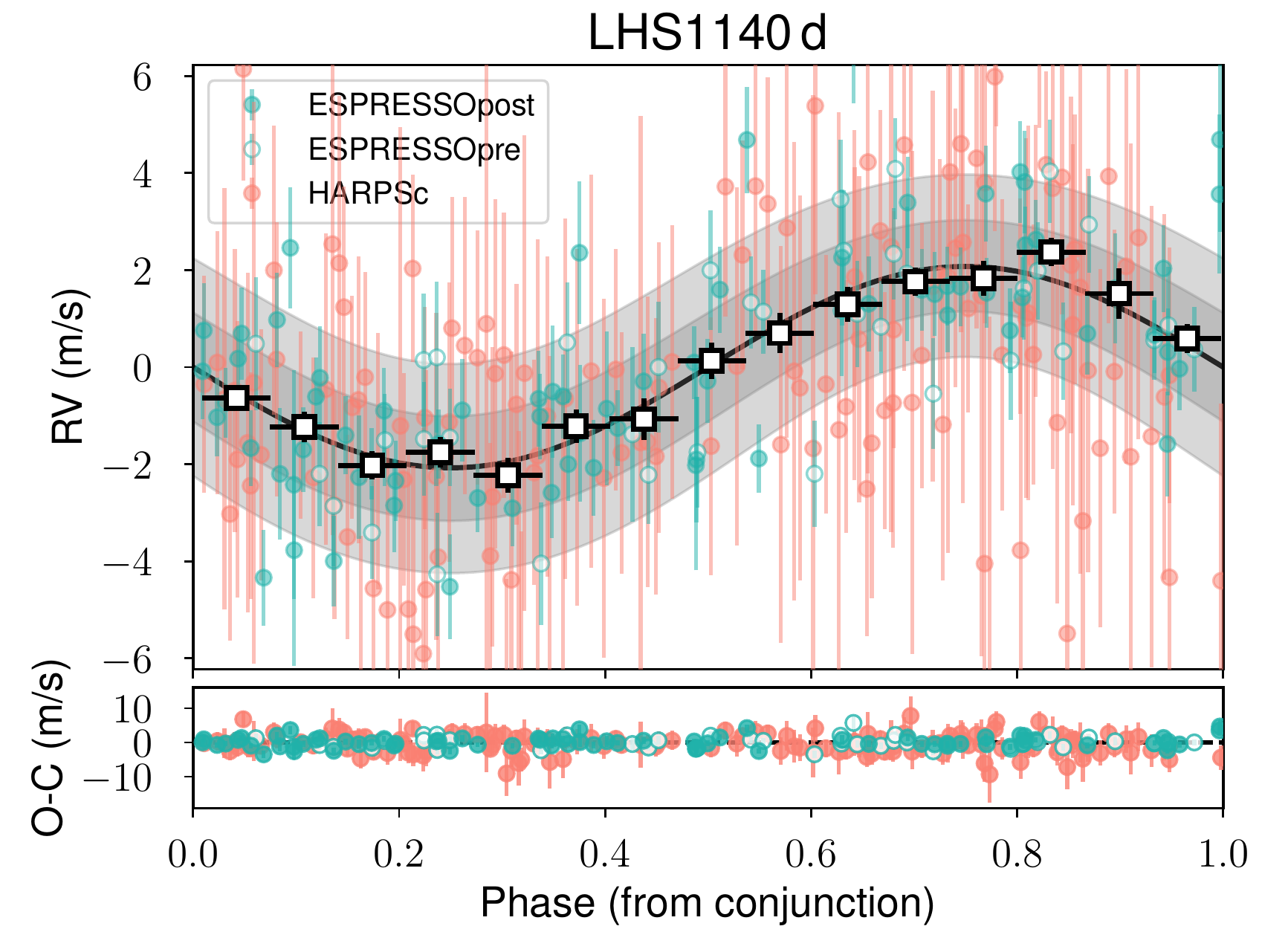}
\caption{Phase-folded radial velocity signal of the candidate planet \lhsd{} in the three-planet model. The upper panel shows the radial velocity data (color-coded as in the legend), binned radial velocities (open black symbols) with a bin size corresponding to 15\% of the phase, the median model for from the joint three-planet analysis (see Sect.~\ref{sec:joint}) as the solid black line, and the 68.7\% and 95\% confidence intervals as shaded dark and light gray regions. The bottom panels show the residuals of the median model.}
\label{fig:lhsd}
\end{figure}

Although the evidence in the full dataset is weaker, all three-planet models in the combined HARPS+ESPRESSO dataset converged to a planet candidate with a period of $\text{about }$80 days, corresponding to the peak in the $\ell 1$-periodogram described in Sect.~\ref{sec:l1per} and already seen in the classical periodogram presented in Sect.~\ref{sec:espressoRVs} and Fig.~\ref{fig:periodograms}. The three-planet model with all planets on circular orbits (labeled 3p1c2c3c) has the second strongest evidence, even above the eccentric two-planet cases. The median and 68.7\% confidence interval of the parameters for this circular three-planet model are presented in the last column of Table.~\ref{tab:rv3p}{, and} the phase-folded radial velocity curve of this third planet candidate is shown in Fig.~\ref{fig:lhsd}. It is important to note that this strong evidence is mainly due to the addition of the exquisite ESPRESSO data, which itself show clear signs for this three-planet model. 

\begin{figure}
\centering
\includegraphics[width=0.5\textwidth]{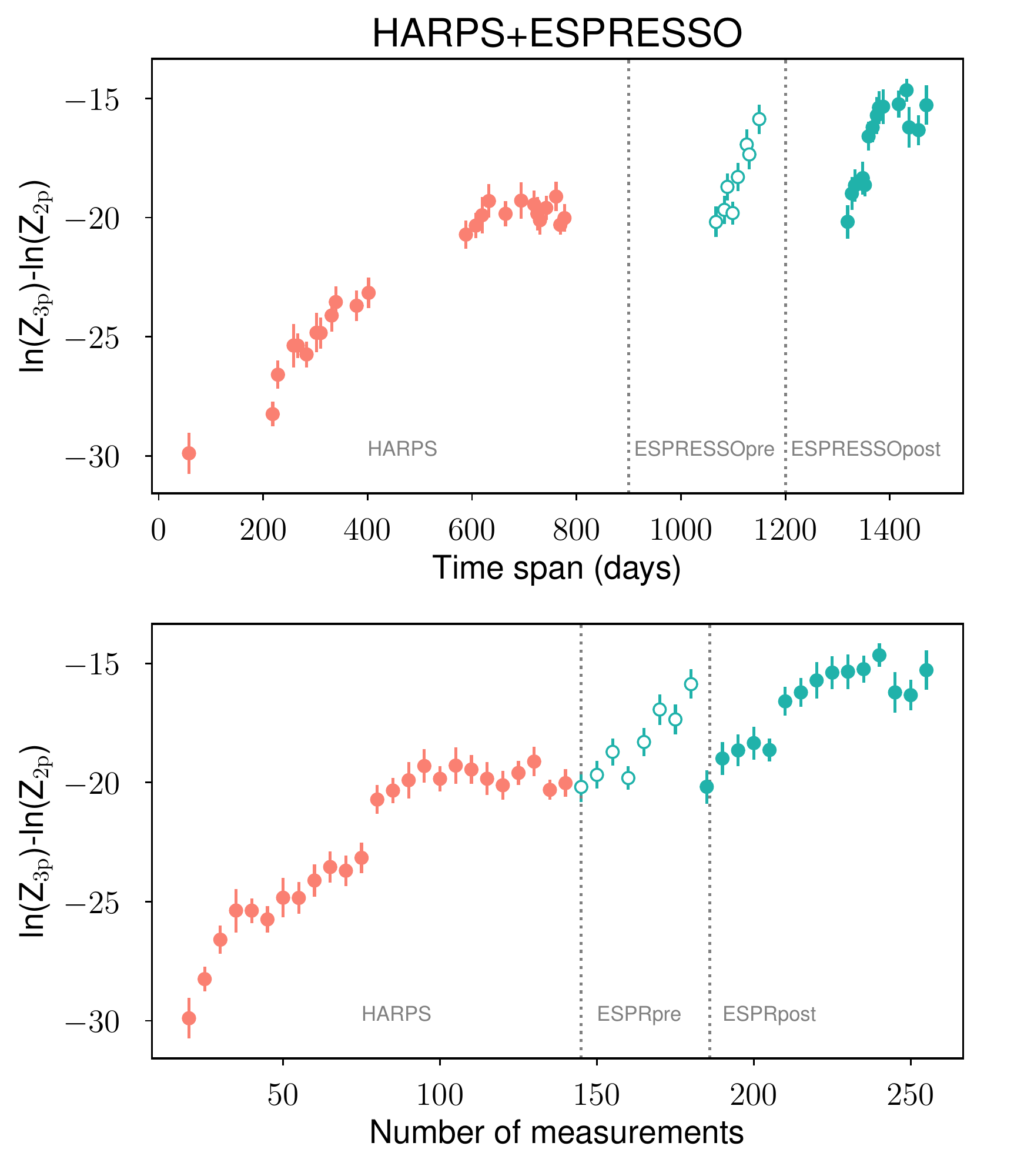}
\caption{Difference in the log-evidence of the two-planet and three-planet models with circular orbits for a cumulative number of data points. The top panel shows the Bayes factor against the time span of the data set, and the bottom panel shows this against the number of data points. }
\label{fig:recursive_evidence}
\end{figure}

We performed a cumulative Bayesian analysis of the dataset by obtaining the Bayes factor for the two most likely models with two and three planets (i.e., the circular case for all planets, labeled 2p1c2c and 3p1c2c3c in Fig.~\ref{fig:evidence}). We obtained the log-evidence using up to $N_i$ data points in steps of five data points and starting with the first 20 measurements. The results are presented in Fig.~\ref{fig:recursive_evidence} and show that the log-evidence of the three-planet model, although the evidence is weaker than for the two-planet model, increases progressively when more data points are added. This is an indication that additional data might clearly confirm this signal. The jumps in the Bayes factor correspond to the addition of new data from a different instrument, which correspondingly adds new parameters (and so complexity) to the model. {The evolution of the Bayesian evidence shows that about 50 new ESPRESSO measurements would be needed to clearly confirm this candidate signal.}

According to our three-planet model in the combined dataset, the third planet would have a minimum mass of $m_d\sin{i_d}=4.8^{+1.3}_{-1.2}$~\Mearth{} and an orbital period of $P_d=79.22^{+0.55}_{-0.58}$ days. This describes a world in the rocky-gaseous frontier beyond the habitable zone. We note here that the detection of its radial velocity signal is at the $\sim3.8\sigma$ level ($2.21^{+0.59}_{-0.57}$). According to the ephemeris found in this radial velocity solution, the TESS observations would have missed the transit of this small planet (in case of a coplanar orbit with the other known planets in the system, the estimated impact parameter is 0.65), which would have occurred about five days before TESS observations started in this sector. The estimated radius of this planet according to \texttt{forecaster} \citep{chen17} is $2.0^{+0.93}_{-0.63}$~\Rearth{} , which places the posterior distribution for the planet radius in the middle of the radius valley. The corresponding transit depth would be about 8 ppt, which is comparable with the transit depth of \lhsb{} and therefore easily detectable by TESS and ground-based instrumentation. TESS will revisit the system in Sector 30 (22 September 2020 to 21 October 2020, in cycle 3). However, given the derived ephemeris for the candidate planet \lhsd{}, the probability of a transit within the TESS observations in Sector 30 is unfortunately only 0.03\% (see the analysis in Sect.~\ref{sec:joint}).

To verify the results, we performed an independent analysis with \texttt{kima} \citep{kima} on the HARPS+ESPRESSO dataset. We considered up to three Keplerian signals in addition to a GP model for the correlated noise. Broad and equal priors were assigned to the three sets of orbital parameters (i.e., we did not use information from the transits). For the orbital periods we used a log-uniform distribution between 1 and 100 days, for the semiamplitudes a uniform prior between 0 and 20 m/s, and for the eccentricities a Kumaraswamy distribution with shape parameters $a=0.867$ and $b=3.03$ \citep{kipping13}. The priors for the GP and remaining parameters were otherwise very similar to those in Table~\ref{tab:rv3p}. 

The results indicate a significant detection of only two signals, corresponding to the orbital periods of the two known transiting planets, and the orbital parameters are fully consistent with those found previously. The probability ratio between the three-planet and two-planet models is estimated at 1.15, leading to the marginal detection of a third signal, with an orbital period of $78.9 \pm 0.5$ days. This is compatible with the results explained above. As before, the data do not constrain the eccentricities in the two-planet model. Consistent results were also obtained for the GP parameters.

%-------------------------------------------------------------------------
\subsection{Limits on additional planets at different periods}
\label{sec:additional}

\begin{figure}
\centering
\includegraphics[width=0.5\textwidth]{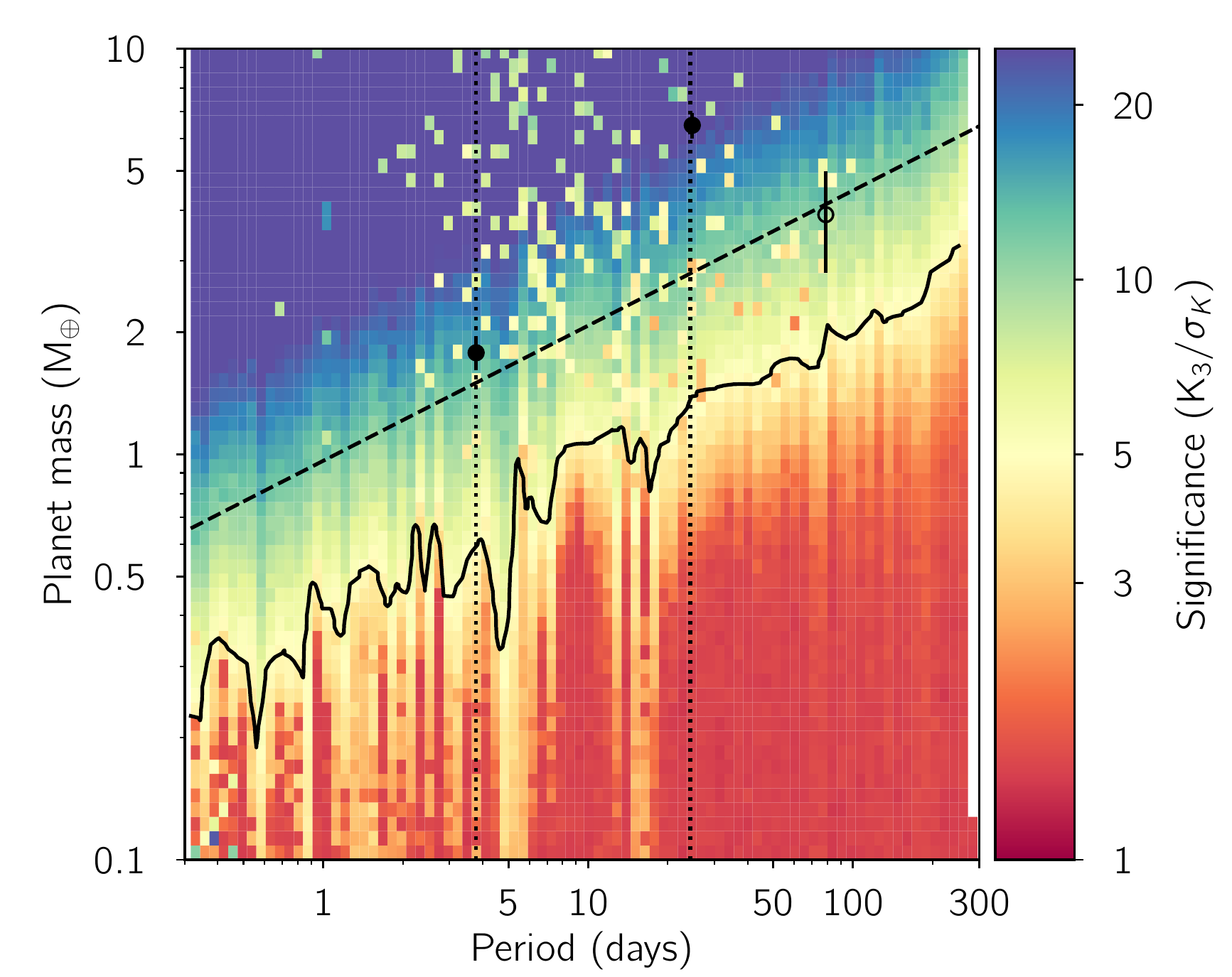}
\caption{Detection limits of the radial velocity dataset for periods shorter than one year. The color code indicates the significance of the radial velocity semiamplitude parameter. The $5\sigma$ contour is shown as a solid line, and the limit corresponding to the scatter of the data is shown as a dashed line. The locations of the two known planets are shown as solid circles, and the location of the third planet candidate is shown as an open circle.}
\label{fig:rvlimits}
\end{figure}

 Because no additional planets can be confirmed with the current dataset, we explored the sensitivity of the HARPS and ESPRESSO data and sampling by injecting planetary signals (we assumed circular orbits) for different periods and masses ranging from 0.3-1000 days and 0.1-300~\Mearth{}. In total, $10^4$ signals were injected in the GP-removed dataset\footnote{We used the GP-removed dataset for computational efficiency reasons because the GP computation with \texttt{george} takes several hours in our MCMC algorithm for just one model. Analyzing 10$^4$ time series would take months of computational time in our HPC cluster.}. Then we proceed in the same manner as described in Section~\ref{sec:rv3p} (but we now removed the GP part) to recover the injected planet signal. We considered the planet to be detected when the median of the posterior distribution for the radial velocity semiamplitude was more than five times the standard deviation away for zero. This analysis is shown in Fig.~\ref{fig:rvlimits}, where we also add the detection limit assuming the scatter of the radial velocity data from the two-planet model in Section~\ref{sec:rv3p}. We conclude that we can discard planets with masses $m>1$~\Mearth{} for periods up to 10 days, $m>2$~\Mearth{} for periods below 100 days, and $m>5$~\Mearth{} for periods shorter than one year. We note that these limits refer to the detection of the planet signal. The statistical significance of the model with respect to the two-planet model is not assessed in this step. The signal from the planet candidate \lhsd{}  is therefore above the sensitivity line in Fig.~\ref{fig:rvlimits} (we can detect it in the data), but it cannot be confirmed with sufficient statistical significance.

%===================================================
\section{Exploring the TESS dataset}
%===================================================
\label{sec:lc_explore}

The full analysis of the transit signal of \lhsb{} and \lhsc{} is presented in Section~\ref{sec:joint}. Here, we explore the possibility of additional signals in the TESS dataset.

We first analyzed the TESS dataset to search for transit-timing variations (TTVs) of \lhsc,{} for which five transits are present in the data. We used \texttt{allesfitter} \citep{gunther20} to model the photometric data alone. The priors on the parameters were set to Gaussian distributions according to the values published in \cite{ment19}, with a broad width corresponding to ten times the published uncertainties. We also set uniform priors to the individual TTVs of each transit of $\pm3.5$~hours. The results for the inferred TTVs are shown in Table~\ref{tab:ttvs}. All values are compatible with zero TTVs, except for the first transit, which occurs slightly earlier. However, the TTV is still compatible with zero within 2$\sigma$. Additionally, the TESS transit times are also compatible within one minute with the Spitzer transit time obtained six months before by \cite{ment19}. 

\begin{figure}
\centering
\includegraphics[width=0.49\textwidth]{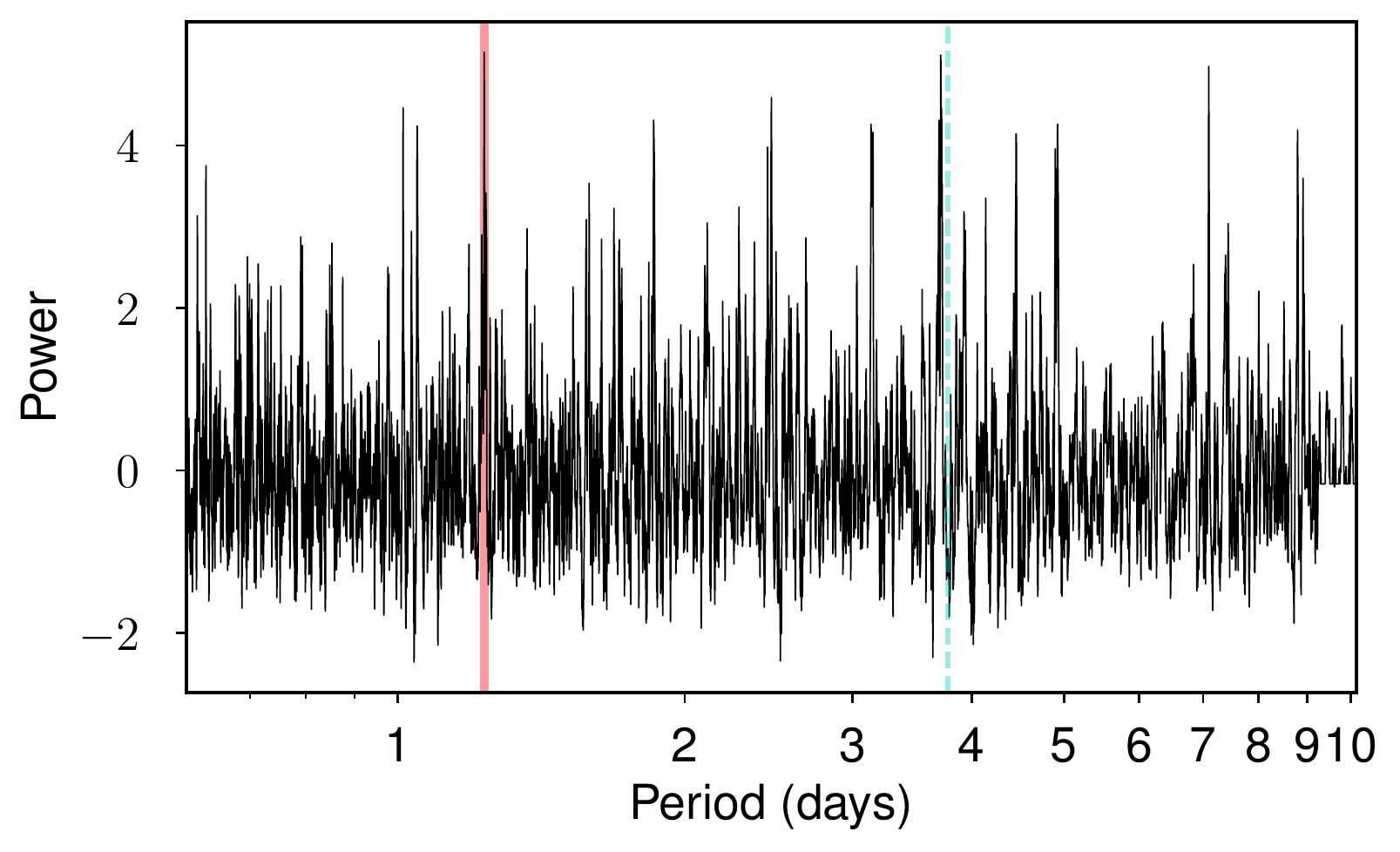} %/Users/lillo_box/00_projects/11__LHS1140/TESS/
\caption{TLS periodogram of the TESS dataset after masking the light curve from the transit times corresponding to the two known planets \lhsb{} and \lhsc{}. The red line indicates the strongest power peak, and the blue line indicates the period of \lhsc{}.}
\label{fig:tls}
\end{figure}

We used the transit least-squares (TLS) software \citep[\texttt{tls}, see][]{heller19} to search for additional signals in the TESS dataset. We masked out the times of transit corresponding to planets \lhsb{} and \lhsc{} and performed a smooth detrending of the light curve using \texttt{w{\={o}}tan} \citep{wotan} with a window length of 0.5 and using the \textit{biweight} method. We then computed the TLS periodogram on this masked and detrended light curve over the period range calculated by the \texttt{tls} software based on the data sampling with a range spanning from 0.6 to10 days. The result is shown in Fig.~\ref{fig:tls}. The highest peak in the TLS periodogram corresponds to a periodicity of 1.26 days at a signal detection efficiency (SDE) of 6. Other intriguing signals are also present: a signal close to the orbital period of \lhsc{}. Although the signal does not reach the statistical significance level of $SDE>7$ required to accept a periodicity as statistically valid (see \citealt{heller19}), the fact that it appears at a similar orbital period as \lhsc{} is intriguing and is further analyzed in Sect.~\ref{sec:coorbitals} and in particular in Section~\ref{sec:coorbitals_c}.

\begin{table}[]
\setlength{\extrarowheight}{3pt}
\caption{\label{tab:ttvs}Priors and posterior distributions for the radial velocity analysis (see Sect.~\ref{sec:rv3p}).}
\begin{tabular}{lcc}
 
 \hline 
{Transit \#} & $T_0$ (BJD) & TTV (min)  \\ 
\hline 
1 & $2458389.28715_{-0.00419}^{+0.00477}$ &  $-9.31_{-5.86}^{+6.72}$\\
2 & $2458393.07194_{-0.00403}^{+0.00364}$ &  $0.56_{-5.61}^{+5.05}$\\
3 & $2458396.85022_{-0.00231}^{+0.0023}$ &  $1.06_{-2.99}^{+3.0}$\\
4 & $2458400.62857_{-0.00234}^{+0.00236}$ &  $1.67_{-3.05}^{+3.09}$\\
5 & $2458404.40569_{-0.00264}^{+0.00277}$ &  $0.5_{-3.51}^{+3.73}$ \\
 \hline
 
 \end{tabular}

\end{table}

%===================================================
\section{Co-orbital analysis}
%===================================================
\label{sec:coorbitals}

The long time span and high precision of the radial velocity dataset obtained for this target allows us to perform a detailed exploration for the possible presence of co-orbital planets (exotrojans). We explored this scenario for both planets in the system by applying the technique described in \cite{leleu17} (hereafter the $\alpha$-test), a generalization of the technique proposed by \cite{ford06}. The $\alpha$-test has been applied to different planetary systems, for instance, by  \cite{lillo-box18a,lillo-box18b}, \cite{armstrong20}, or  \cite{toledo-padron20}. In this technique, the radial velocity induced by the co-orbital system corresponds the the sum of two Keplerians with the same orbital period, with at least one of the components transiting its host star (assumed to be the planet). When we also assume that the mass of the planet pair ($m_p+m_t$) is far lower than the stellar mass ($M_{\star}$) and that in the near-circular case (typically, $e<0.1$), the radial velocity can be approximated to first order in eccentricity and mass by 
\begin{equation}
\begin{aligned}
v (t) =  \gamma + \mathcal{K} \big[&(\alpha -2c) \cos n t  - \sin n t \\
       &+ c \cos 2 n t + d \sin 2 n t \big] \, ,
\end{aligned}  
   \label{eq:RVf}
   \end{equation}
\noindent where $n=2\pi/P_{\rm orb}$, $\gamma$ is the systemic velocity, and $\mathcal{K}$ is the radial velocity semiamplitude of the co-orbital pair. The most relevant parameter in this equation is $\alpha$, which to first order in eccentricity and trojan-to-planet mass ratio is
\begin{equation}
\begin{aligned}
    \alpha & = - \frac{m_t}{m_p} \sin \zeta + \gO\left[\left(\frac{m_t}{m_p}\right)^2,e_k^2,\frac{m_t}{m_p} e_k\right]\, ,
\end{aligned}  
 \label{eq:alphalim}
   \end{equation}
\noindent where $\zeta$ is the difference in the mean longitude of the two components of the co-orbital pair, and $t$ is the time, with the origin set to the mid-transit time of the main planet. This $\alpha$ parameter is therefore zero when no co-orbital is present (as $m_t=0$) and can be qualitatively described as the mass ratio between the trojan and the planet. If the radial velocity data are compatible with a nonzero value for this parameter, we can infer a potential mass imbalance between the planet and the locations of the Lagrangian points. A negative value corresponds to \lfour{} and a positive value to \lfive{}.

%-------------------------------------------------------------------------
\subsection{Co-orbital analysis for \lhsb{}}
\label{sec:coorbitals_b}

The main interest in this planet resides on its location (in the middle of the habitable zone of this M dwarf) and its physical properties, including a rocky composition and low eccentricity (if any). These properties make this planet ideal for co-orbital searches and add the interest that it might lie in the habitable zone. \cite{dvorak04} have demonstrated that rocky trojans that co-orbit with planets in the habitable zone can also be habitable.

According to the results presented in Section~\ref{sec:rv3p}, we here assumed only planets \lhsb{} and \lhsc{} in the system. We therefore  added another Keplerian signal to Eq.~\ref{eq:RVf} to account for the radial velocity contribution of \lhsc{}. We followed the same modeling procedure as in Section~\ref{sec:rv3p}. We again tested all possible scenarios corresponding to the two-planet models in that section (i.e., we assumed circular or eccentric orbits for the two planets).

We find the strongest evidence for the circular model for both planets \lhsb{} and \lhsc{}. In this model the inferred $\alpha$ parameter is $\alpha=0.015\pm0.065$. The estimated value is fully compatible with zero, clearly stating that no co-orbital is detectable down to our sensitivity limits. Instead, using the posterior distribution of this parameter, we can set limits to the presence of co-orbitals to this planet at both Lagrangian points. Assuming the 95\% confidence level as an upper limit, we can discard co-orbital planets to \lhsb{} more massive than 1.8~\Mearth{} at \lfour{} and 1.0~\Mearth{} at \lfive{}. Using only the HARPS measurements, we obtained a broader distribution of $\alpha=0.068^{+0.097}_{-0.096}$, which provided upper limits of 2.1~\Mearth{} at \lfour{} and 1.8~\Mearth{} at \lfive{}. By adding the 116 ESPRESSO data points, we can decrease this limits by $>30$\%.

We also inspected the TESS light curve around the location of the Lagrangian points \lflf{} of \lhsb{}. Unfortunately, the \lfour{} location falls into a gap in the middle of the TESS sector. The the transit of \lfive{} region occurs at $\rm{BTJD}=1404.05$, about eight hours before a transit of the inner planet \lhsc{}. At the exact location of the Lagrangian point passage, no transit is found within the photometric precision. The scatter of this region in the light curve is 2.1 ppt (parts per thousand), which would correspond to a $3\sigma$ upper limit on any object corresponding to 3~\rearth.

%-------------------------------------------------------------------------
\subsection{Co-orbital analysis for \lhsc{}}
\label{sec:coorbitals_c}

We proceeded in the same manner as for \lhsb{}, but now included the $\alpha$ parameter for planet c. In this case, we obtain a value of $\alpha=-0.129^{+0.087}_{-0.090}$. The 95\% confidence interval is between $\alpha \in [-0.27, 0.02]$. This means that the 95\% upper limit (corresponding to $2\sigma$) is only marginally compatible with zero. A further exploration is therefore valuable. Because of its negative value, the mass imbalance producing this radial velocity signal can be interpreted as an existing body located at \lfour{}. Assuming the mass of \lhsc{} as calculated in Sect.~\ref{sec:rv3p} and a coplanar orbit, the mass of the trojan\footnote{This value assumes $\zeta=60^{\circ}$. It therefore represents an average status of the system if the time span of the observations is longer than the libration period, or an instantaneous mass otherwise.} would correspond to $0.26\pm0.18$~\Mearth. Despite this $\sim 2\sigma$ signal, the Bayesian evidence of the co-orbital model is still lower than the two-planet model. For the HARPS-only dataset, the evidence is stable throughout the entire dataset. However, in the case of the ESPRESSO-only dataset, the evidence for the co-orbital model progressively (although slowly) increases toward positive values as more data points are added. It therefore deserves additional attention.

We also performed an independent test to confirm this signal by following the technique proposed by \cite{ford06} to compare the time of conjunction measured from the transit signal ($T0_{\rm c,LC}$) and from the radial velocity signal ($T0_{\rm c,RV}$). In the case of a single planet in circular orbit, it is  easy to see that the lag between both times is zero, $\Delta t = T0_{\rm c,LC} - T0_{\rm c,RV} = 0$. However, in the presence of a co-orbital body at one of the Lagrangian points, the gravitational pull of the co-orbital will change $T0_{\rm c,RV}$ (this occurs earlier if in \lfour{} and later if in \lfive{}), while  leaving the time of transit ($T0_{\rm c,LC}$) unchanged. We performed this test by analyzing both datasets independently. We modeled the TESS light curve with two planets and the radial velocity dataset with two Keplerian signals (details of the modeling of the photometric dataset are provided in Sect.~\ref{sec:joint}). As a result, we obtain a time lag between the times of conjunction for \lhsc{} measured from both techniques of $\Delta t = 2.04 \pm 1.27$~hours. This value is again ~1.6$\sigma$ away from zero and provides consistent results with the $\alpha$-test, an indication for a co-orbital body in \lfour{} of \lhsc{}. We note that this time lag can also be explained by a single planet with $e\cos{\omega} \sim 0.15$, which is within the confidence interval of the eccentric two-planet model we showed in Sect.~\ref{sec:rv3p}.

\begin{figure}
\centering
\includegraphics[width=0.49\textwidth]{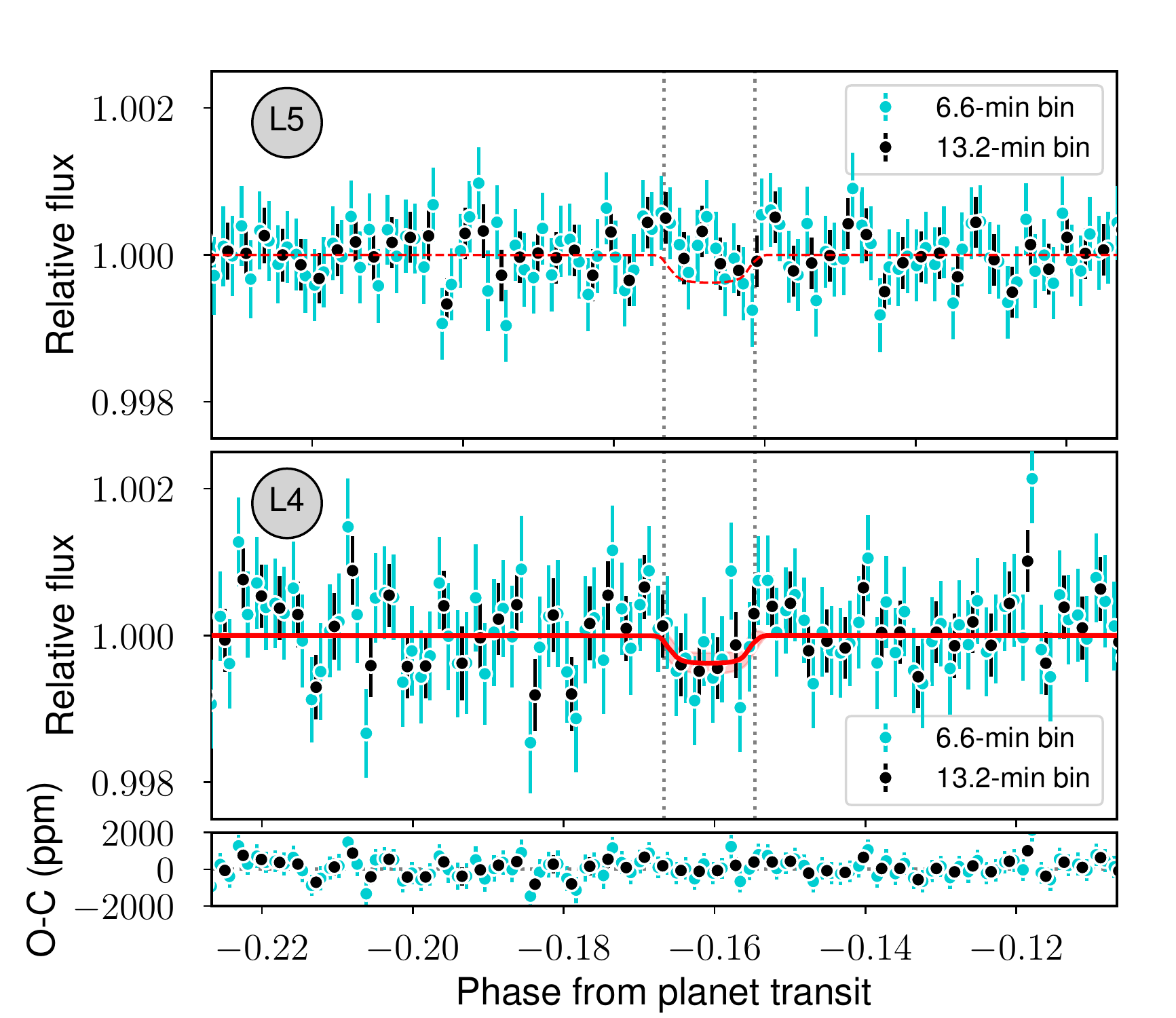}
\caption{{TESS phase-folded light curve around the Lagrangian points \lfive{} (top panel) and \lfour{} (middle panel) of \lhsc{} (see Sect.~\ref{sec:coorbitals_c}). The expected location of the transit is marked by vertical dashed lines. The median transit model inferred from the analysis of the \lfour{} dimming is shown as a solid red line (middle panel) and a dashed line in the upper panel to guide the eye on the photometric scatter of the \lfive{} region. The $1\sigma$ confidence interval is shown in the middle panel as a shaded red region. The bottom panel shows the residuals of the model for the \lfour{} region. All panels show bin data points corresponding to 6.6 minutes (light blue, ten data points inside the transit duration) and 13.2 minutes (black, five data points inside the transit duration).}}
\label{fig:coorbital_tess}
\end{figure}

We further investigated this co-orbital candidate by inspecting the TESS light curve in this Lagrangian point. Five transits are observed with TESS of \lhsc,{} but only four passages of its \lfour{} in front of the star because the third Lagrangian point lies inside the mid-sector gap. We explored the TESS light curve in the region of the Lagrangian point after removing the data points inside the transit of planet \lhsb{} which occurs close to the \lfour{} location of transit 4 of \lhsc{}. The phase-folded light curve in this regime shows a shallow dimming at the location of the \lfour{} Lagrangian point (slightly shifted by 0.006 in phase, or equivalently, $\sim$32 minutes, but perfectly compatible with a dynamically stable libration orbit), see the middle panel in Fig.~\ref{fig:coorbital_tess}. Interestingly, the duration of the dimming is perfectly compatible with the transit duration of the planet \lhsc{}, expected for a co-orbital in the coplanar case. {The depth of this dimming is at the same level ($\sim 370$~ppm) as the photometric scatter in the out-of-transit regions, however. We can compare this with the \lfive{} region. The photometric scatter in \lfive{} is significantly smaller ($\sim 280$~ppm), and no signs of dimming are seen at the \lfour{} depth. This may indicate that the origin of the dimming is related to the larger systematics around the \lfour{} region. However, the different indications (including the radial velocity offset, the duration of the dimming corresponding to the planet transit duration, and its location close to the Lagrangian point) encouraged us to further investigate this signal. } We therefore modeled it together with the signal of the two known planets to retrieve the size of the potential co-orbital candidate. The result provides a radius for this candidate of $0.44\pm0.08$~\Rearth. This result increases the certainty in the co-orbital scenario. However, the shallow depth {(compatible with the photometric noise in the out-of-transit regime around \lfour{})} prevents us from definitively confirming the nature of this dimming. If confirmed, it might also explain the peak in the TLS periodogram close to the period of \lhsc{} described in Sect.~\ref{sec:lc_explore} (see also Fig.~\ref{fig:tls}). Additional data from the TESS extended mission will shed more light on this signal.

\begin{figure*}
\centering
\includegraphics[width=1\textwidth]{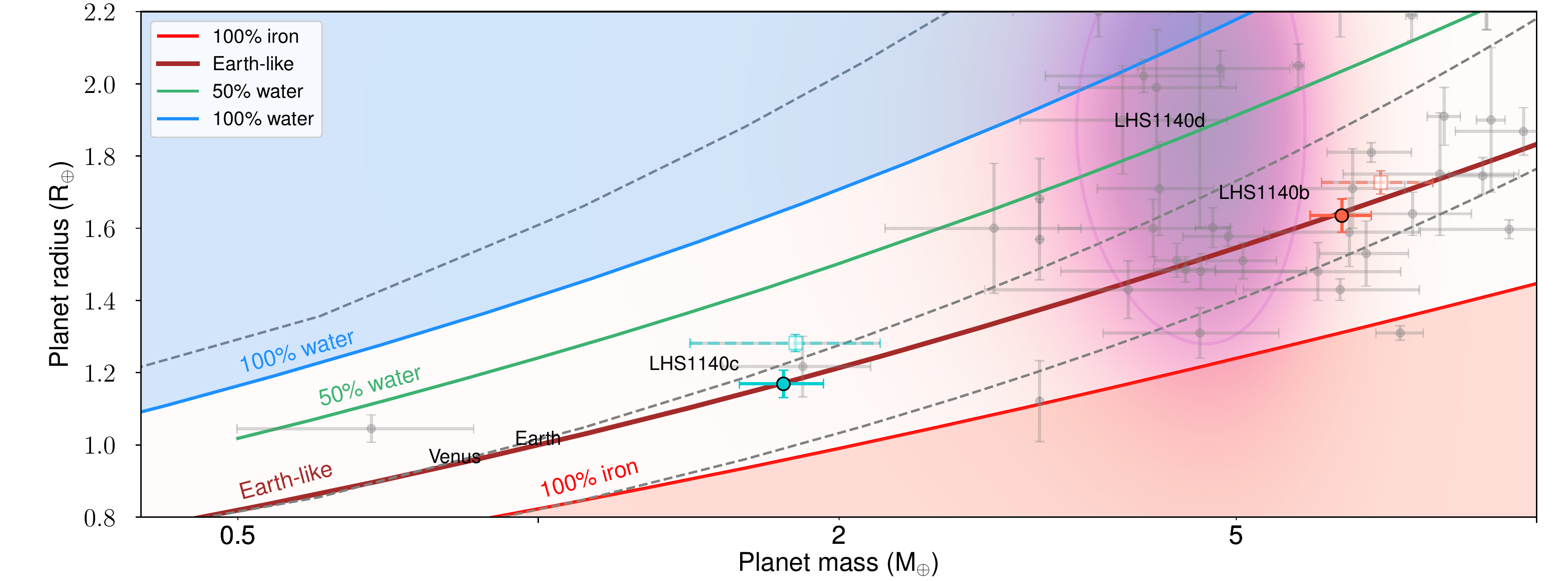}
\caption{Mass-radius diagram for the known exoplanets with the lowest masses with measured mass precisions better than 30\% (gray symbols). The new locations of the \lhs{} planets are shown as filled blue (\lhsb{}) and orange (\lhsc{}) circles. Their previous locations using the values from \cite{ment19} are shown as open squares with dashed error bars and the same color code. The candidate planet at an orbital period of $\sim$78 days is shown as the shaded magenta region. Its radius is not known, therefore its vertical location comes from the simple mass-radius relations from \cite{chen17}. The bulk density lines corresponding to different compositions from \cite{zeng19} are shown as solid traces, and the dashed lines correspond to iso-densities of 1.33, 5.3, and 10 g$\cdot$cm$^{-3}$ (from top to bottom).}
\label{fig:mr}
\end{figure*}

%===================================================
\section{Joint photometric and radial velocity model}
%===================================================
\label{sec:joint}

Based on the analysis performed in the previous section, we proceeded with a joint simultaneous analysis of the HARPS and ESPRESSO radial velocities and the TESS photometric time series to estimate the physical and orbital parameters of the system. To this end, we modeled the the radial velocities as explained in Sect.~\ref{sec:rv3p}, but with a slightly different parameterization that now explicitly included the individual contributions of the planet mass and orbital inclination for each planet (instead of using the radial velocity semiamplitude). We also included the GPs in the modeling and constrained them by simultaneously modeling the FWHM of the CCF of the ESPRESSO and HARPS data (see Sect.~\ref{sec:rv3p}). 

The TESS photometry was modeled with the \texttt{batman} code \citep{kreidberg16} to retrieve the transit models. We assumed a quadratic limb-darkening model with Gaussian priors around the values calculated by using the \texttt{limb-darkening} code\footnote{\url{https://github.com/nespinoza/limb-darkening}} \citep{espinoza15} for the stellar parameters of \lhs{} published in \cite{ment19}. We also added photometric jitter to account for underestimated white noise, a mean level parameter to account for imperfect normalization of the light curve, and a dilution factor to account for the contamination in the TESS light curve due to the additional sources in the aperture (see Fig.~\ref{fig:tpf}). The stellar mass and radius were also included in the fit to properly account for their uncertainties by using a Gaussian prior around the published values by \cite{ment19}.

To sample the posterior distributions, we followed the same principles as in Sect.~\ref{sec:rv3p}. We used an MCMC sampler (\texttt{emcee}) with 112 walkers (four times the number of parameters) and 30\,000 steps per walker. We finally removed the first half of each chain and combined all chains ($1.68\times10^6$ steps in total) to compute the final posterior distributions. Table~\ref{tab:joint} shows the prior and posterior distributions for each parameter. The final models and confidence intervals for the planet transits are shown in Fig.~\ref{fig:tess}, and the corresponding modeling of the radial velocity dataset is shown in Fig.~\ref{fig:rvtime} (full time series) and Fig.~\ref{fig:rvphase} (phase-folded dataset). In Fig.~\ref{fig:fwhm} we additionally show the joint modeling of the FWHM of the CCF sampling the stellar activity. The variations seen in this parameter mimic the variations seen in the radial velocity dataset, showing the correspondence between both measurements coupled by the activity of the star (see also \citealt{suarez-mascareno20} for a similar analysis on Proxima\,b). 

The inferred masses are compatible with the minimum masses obtained in the radial velocity analysis (Sect.~\ref{sec:rv3p}). We find masses for the two planets that correspond to $m_{\rm b}=6.38\pm0.45$~\Mearth{} and $m_{\rm c}=1.76\pm0.17$~\Mearth{}, slightly lower than the previously reported values. We also find slightly smaller radii for the two planets with the TESS data ($R_{\rm b}=1.635\pm0.46$~\Rearth{}, $R_{\rm c}=1.169\pm0.038$~\Rearth{}) than the values found by \cite{ment19} using Spitzer data  ($R_{\rm b}=1.727^{+0.032}_{-0.032}$~\Rearth{}, $R_{\rm c}=1.282\pm0.024$~\Rearth{}); they are different by about 1.5$\sigma$  for \lhsb{} and by 2$\sigma$  for \lhsc{}.  

We also performed a three-planet modeling of the full radial velocity and TESS dataset to extract better constraints on the ephemeris of the third planet candidate. The resulting phase-folded radial velocity curve of the third planet is shown in Fig.~\ref{fig:lhsd}. We can better constrain the time of conjunction for this planet to $T_{\rm 0,d} = 2458381.8_{-3.2}^{+2.5}$.

%===================================================
\section{Discussion}
%===================================================
\label{sec:discussion}

%-------------------------------------------------------------------------
\subsection{Mass-radius diagram}
\label{sec:mr}

We estimated the physical and orbital properties of the two known planets with high precision thanks to the precise radial velocity measurements from ESPRESSO and the high-cadence and high-precision photometry from TESS. In Fig.~\ref{fig:mr} we show a mass-radius diagram including all planets with masses lower than 10~\Mearth{} and radii smaller than 2.2~\rearth{} . The mass and radius precisions are better than 30\%. We include the two confirmed planets in the \lhs{} system with their derived properties estimated in this work and the previously known properties from \cite{ment19}. The newly derived properties of both planets place them at the top of the Earth-like bulk density line, while previous measurement suggested a slightly lower density for these planets. The planet radius of \lhsb\  places it close to the radius gap \citep{fulton17}, and the newly derived mass confirms the rocky nature of the planet. \lhsc{} is one of the few planets with a derived mass lower than 2~\Mearth{} in this diagram. The other two planets are GJ\,357\,b \citep{luque19,jenkins19} and TRAPPIST-1\,f \citep{gillon16}. 

%-------------------------------------------------------------------------
\subsection{Transit of the planet candidate \lhsd{}}
\label{sec:mr}

\begin{figure}
\centering
\includegraphics[width=0.5\textwidth]{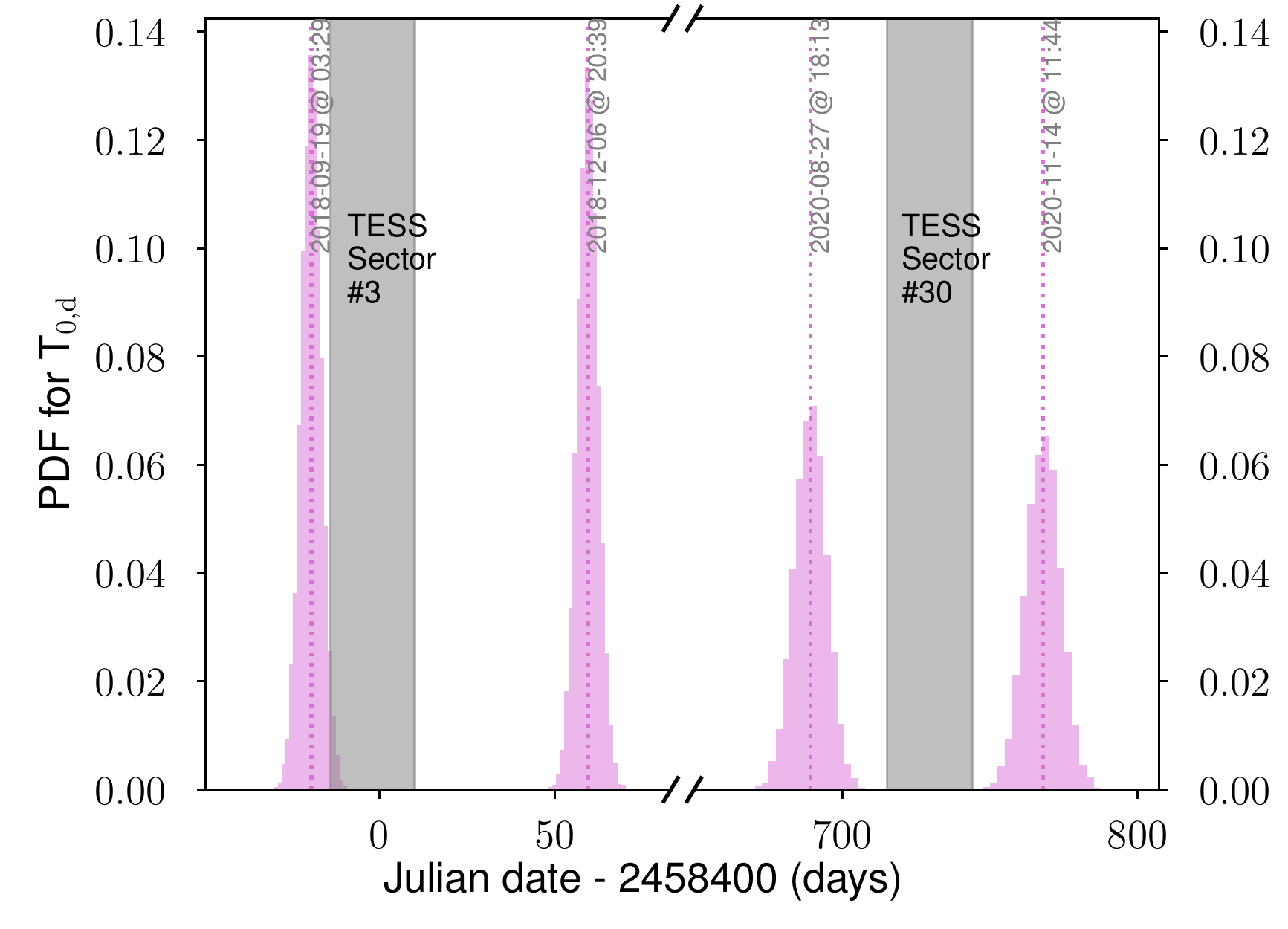}
\caption{Expected posterior distribution of the mid-transit time of planet candidate \lhsd{} (magenta histograms) and the TESS coverage of Sectors 3 and 30 on the field including the host star (shaded gray region). The median of each posterior distribution is marked by a vertical dotted line, and the corresponding calendar date in Universal Time Coordinate system is annotated.}
\label{fig:tess_vissibility}
\end{figure}

The mass-radius diagram in Fig.~\ref{fig:mr} also shows the location of the third planet candidate ($P_d\sim 78$ days) in the system, with an estimated mass of 3.9~\Mearth{}. The uncertainty on the planet radius is taken from the \texttt{forecaster} estimation. The measured mass and estimated radius do not allow us to discern between rocky or gaseous compositions. Based on its mass, the expected planet radius might be in the range 1.2-2.6\rearth. This implies a potential transit depth of 3.5 ppt, which is suitable for ground-based instrumentation. Unfortunately, the precision of the time of conjunction from the radial velocity analysis is only about 3.5 days, which means that the ephemeris is too uncertain to plan ground-based observations. The next transits of this planet candidate will occur on Julian dates { $2459167.8 \pm 5.9$ (2020-11-14 $\pm$ 5.9 days), $2459246.5 \pm 6.4$ (2021-02-01 $\pm$ 6.4 days), $2459325.2 \pm 6.8$ (2021-04-20 $\pm$ 6.8 days), $2459403.9 \pm 7.3$ (2021-07-08 $\pm$ 7.3 days), and $2459482.7 \pm 7.6$ (2021-09-25 $\pm$ 7.6 days)}. When a circular orbit coplanar with the other two known transiting planets is assumed, the impact parameter of this planet candidate would be $b/R_{\star}\sim0.65,$ and it is therefore expected to transit its host star. \lhs{} will be reobserved by TESS in its extended mission in Sector 30. However, despite the large uncertainty on the time of mid-transit, the TESS observations fall in between two transits of the planet and will therefore most likely not be detected by TESS (see Fig.~\ref{fig:tess_vissibility}). As a consequence, ground-based observations, and especially high-precision photometry, are needed to {confirm} this planet and further characterize its properties. {Alternatively, \lhs{} is observable with the Cheops space-based telescope \citep{broeg13}, although the large uncertainty on the ephemeris of the transit would require a large effort from the mission.} 

%-------------------------------------------------------------------------
\subsection{Internal structures of \lhsb{} and \lhsc{}}
\label{sec:internal}

\begin{figure}
\centering
\includegraphics[width=0.49\textwidth]{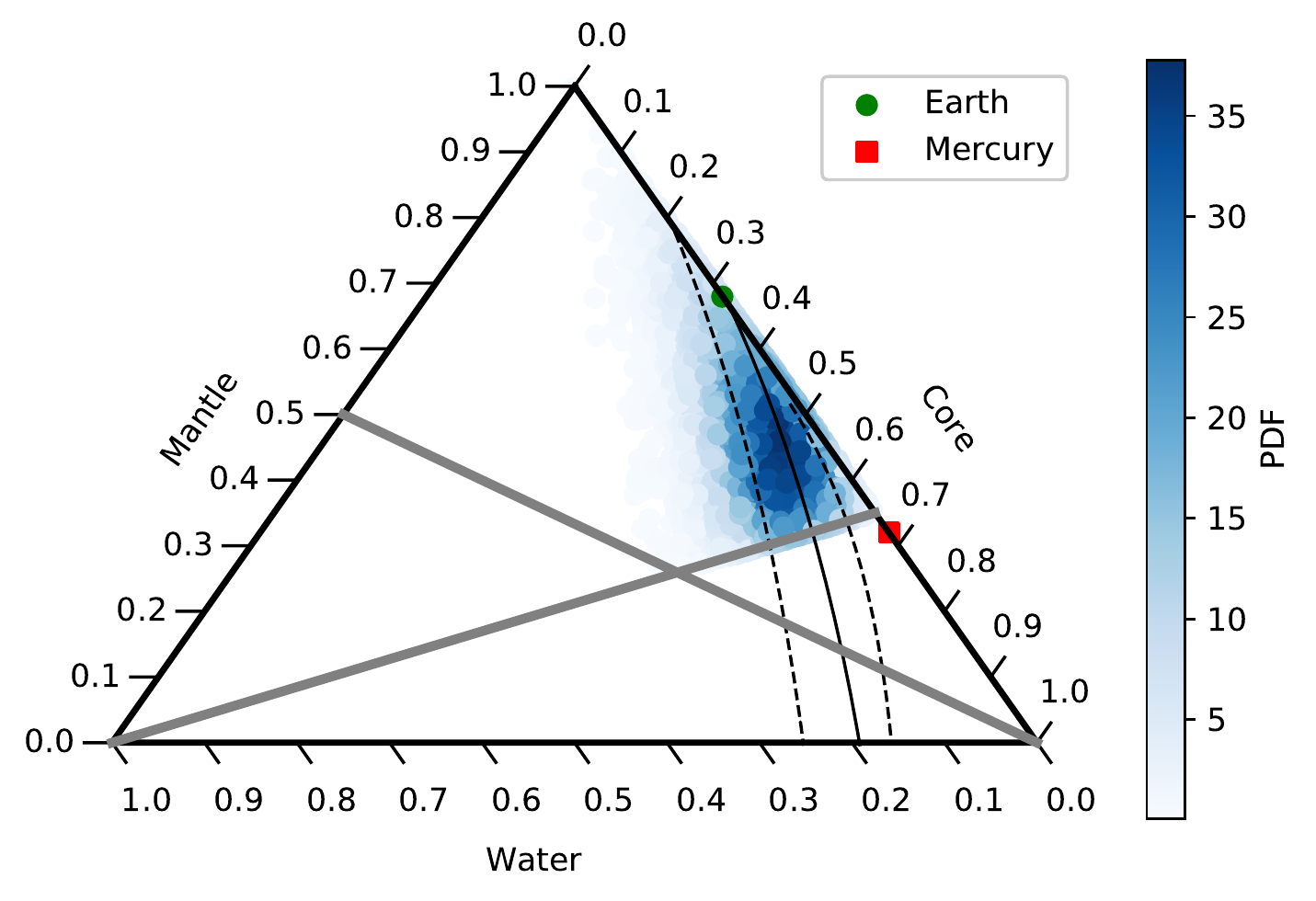}
\caption{Sampled 2D marginal posterior distribution for the CMF and WMF of LHS1140 b. the color code displays the probability density function (PDF). The solid and dashed black lines represent the isoradius curves for the central value of the radius and its 1$\sigma$ confidence interval limits, respectively. The gray lines delimit the areas we excluded from our sampling (see text).}
\label{fig:ternaries}
\end{figure}

\begin{figure}
\centering
\includegraphics[width=0.49\textwidth]{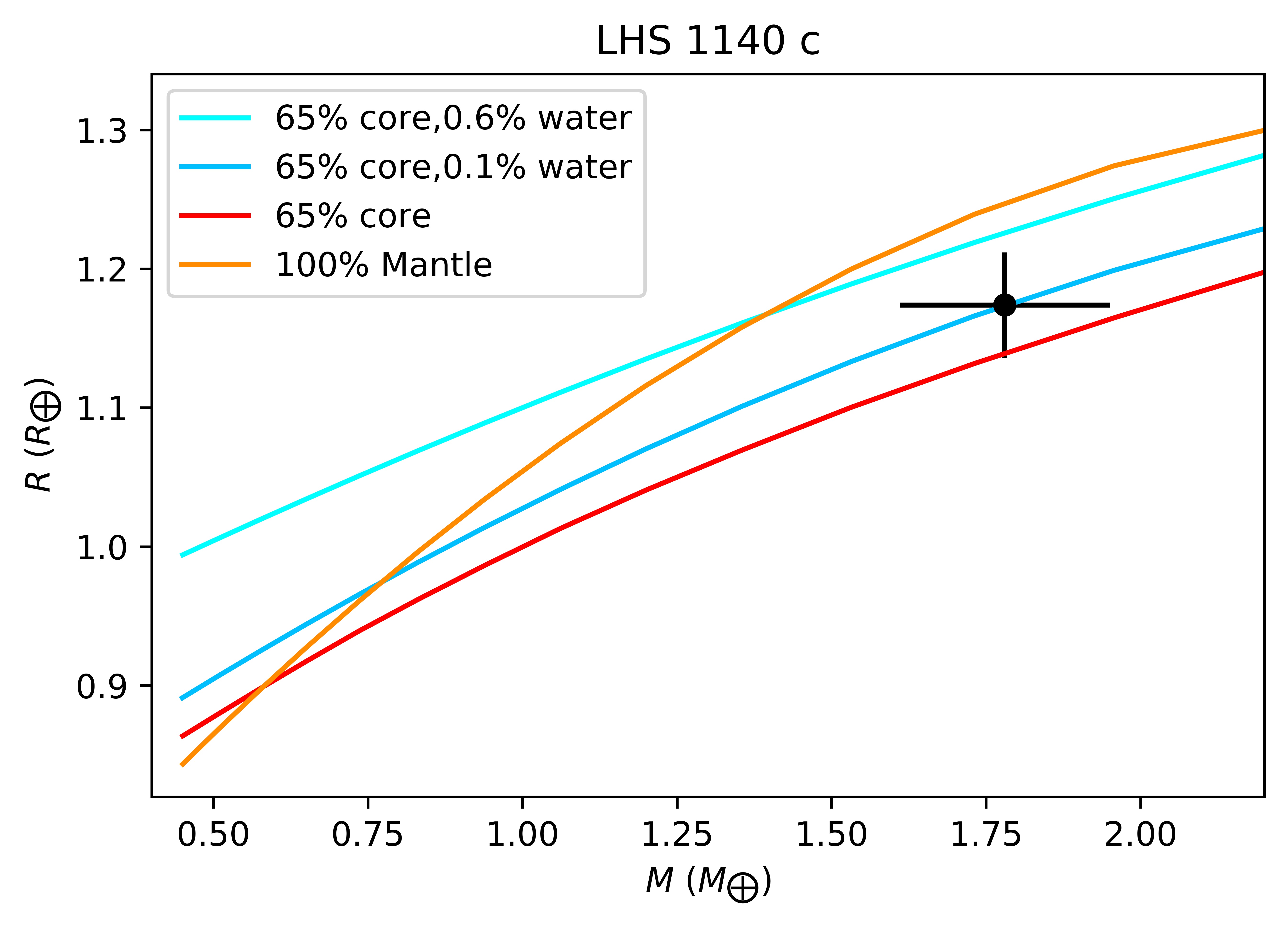}
\caption{Mass-radius diagram for different compositions considering water in supercritical phase under the surface temperature and pressure conditions of \lhsc{}. The black dot and error bars indicate the position of \lhsc{} in the mass-radius diagram.}
\label{fig:MRdiag}
\end{figure}

To investigate the interior of the confirmed planets, we used the internal structure model developed by \cite{Brugger17} for the study of terrestrial planets. The input variables of the interior structure model are the total planetary mass, the core mass fraction (CMF), and the water mass fraction (WMF). In order to explore the parameter space, we performed a complete Bayesian analysis to obtain the probability density distributions of the parameters. The Bayesian analysis was carried out by implementing an MCMC algorithm, following the method proposed by \cite{dorn15}.  The initial values of the three input parameters were randomly drawn from their prior distributions, and we used a Gaussian distribution for the mass and two different uniform distributions for the CMF and the WMF. The uniform distribution for the CMF spans from 0 to 0.65, which is a constraint derived from the Fe/Si ratio in the protoSun \citep{Lodders09}. Including this upper limit, we assumed that these planets have not undergone dramatic processes during or after their formation, such as mantle evaporation or giant impacts. The WMF ranges from 0 to 0.5, the upper value being derived from the composition of the Saturn moon Titan, which is one of the most hydrated Solar System bodies \citep{Tobie06}. Along with these constraints, we took into account the masses and the radii obtained in this study for LHS1140 b and c to derive the posterior probability distributions of their CMF and WMF. We assumed a surface pressure and temperature equal to those prevalent on Earth's surface. This is valid for \lhsb,{} but we note that \lhsc{} is closer to the star, and the WMF estimated in this study therefore only represents an upper limit to its water content (which assumes a layer of liquid water, which is not viable in this case). Furthermore, for  LHS 1140 c, we extended the interior structure model to include the modeling of water in supercritical phase, whose implementation is described in \cite{Mousis20}. We considered an atmosphere with a composition of 97\% water and 3\% carbon dioxide, and a surface pressure at its bottom of 300 bar. The atmospheric mass, thickness, albedo, and surface temperature are provided by a grid generated by the atmospheric model described in \cite{marcq17}.\\

Fig.~\ref{fig:ternaries} shows the results of this internal structure analysis by means of ternary diagrams for both planets. In the case of LHS1140 b, the CMF and WMF are estimated to be 0.49$\pm$0.07 and 0.03$\pm$0.07, respectively. In the case of LHS1140 c, assuming a liquid water layer, the CMF and WMF are estimated to be 0.45$\pm$0.10 and 0.05$\pm$0.07, respectively. This implies that the bulk composition of planet b is expected to be at least 10\% richer in Fe than the Earth's bulk composition. Planet c might be compatible with a CMF of 0.35 within its 1$\sigma$ confidence interval, which is very close to the Earth's 0.32 value. Furthermore, the WMF of \lhsb{} implies a range of water content from no liquid water to 100 times more water than Earth. The posterior distribution peaks at WMF$=0.04$, around 80 times the water content on Earth (0.0005 - 0.5\%- \citealt{sotin07}). The 1\% confidence level corresponds to WMF$_{1\%}=0.007$, still 1.5 times higher than on Earth. When the supercritical phase of water is considered for \lhsc{}, the CMF is calculated as 0.59$\pm$0.05, and the WMF is constrained as $0.0^{+0.006}_{-0.000}$. If we were to assume a CMF similar to that of Earth for \lhsc{}, its WMF would be lower than 0.05\%, which means that the inner planet is very poor in water (see Fig.~\ref{fig:MRdiag}).

%===================================================
\section{Conclusions}
%===================================================
\label{sec:conclusions}

We have revisited the planetary system \lhs{} with its two known Earth-like transiting planets by analyzing an intensive campaign taken with the new ultra-stable high-resolution spectrograph ESPRESSO and observations from Sector 3 of the TESS mission. We have searched for additional signals in the radial velocity data including HARPS and ESPRESSO measurements. The results show additional evidence for the signal of a third planet candidate at $\sim$78 days. Our analysis of the joint radial velocity dataset, including a modeling of the stellar activity using the FWHM of the CCF as an activity indicator, shows positive evidence that this signal isof planetary nature (i.e., not caused by activity). We find a $\sim 4\sigma$ significance in the semiamplitude of the radial velocity of this signal. This is supported by our statistical analysis with \texttt{kima} and the l1-periodogram (see Sect.~\ref{sec:rv3p}). The Bayesian evidence of the two-planet model against the three-planet model, however, is still not enough to claim a confirmation, but its evolution when the dataset was increased points toward the planetary scenario. This evolution shows that $\sim$50 more ESPRESSO measurements are required to unambiguously confirm the planet signal based on radial velocity data alone. The corresponding planet mass of the candidate planet is $3.9\pm1.1$~\Mearth{}. Unfortunately, no transit signal is present in Sector 3 of the TESS data, as expected from the ephemeris of our joint LC+RV analysis. This ephemeris also suggest that the extended TESS mission will not catch the planet transit when the mission revisits this field in Sector 30 between 22 September 2020 and 21 October 2020. If the system is coplanar (as the two known planet inclinations strongly suggest), the impact parameter of this planet candidate would certainly mean that it transits the host star. The expected total transit duration (assuming circular orbit) is $\text{about three}$~hours. Further ground-based photometry to search for this transit would therefore be extremely useful and efficient to confirm this signal. We caution about the relatively large uncertainties of the ephemeris, however, which would imply a dedicated campaign of $\text{about five}$ consecutive nights on the same target. 

We also explored the possibility of co-orbital planets in the two known transiting planets through a dedicated study of the radial velocity data and light curve. For \lhsb{}, the radial velocity data allowed us to constrain the mass of any tadpole and horseshoe co-orbitals up to 2.1~\Mearth{} in \lfour{} and 1.8~\Mearth{} in \lfive{} (95\% confidence levels). Similarly, the TESS data discard nearly coplanar co-orbitals around the Lagrangian point regions larger than $\sim$3~\rearth{}. On the other hand, for \lhsc{}, the radial velocity analysis unveils a 2$\sigma$ signal in its \lfour{} region that would correspond to a sub-Earth co-orbital planet with a 0.26$\pm$0.18~\Mearth. The Bayesian evidence of this model still does not support this scenario against the null hypothesis, {however, and several hundred additional measurements with ESPRESSO-like precision are required to confirm the signal with radial velocity data alone}. However, the analysis of the TESS light curve indicates a shallow dimming very close to this Lagrangian point with the same duration as the transit of \lhsc{} (as expected in the coplanar case). If real, this dimming would correspond to a $0.44\pm0.08$~\rearth{} sub-Earth size body. Unfortunately, the significance of the signal is not strong enough to ensure that this dimming is not caused by systematics of the TESS light curve, and only additional data will shed more light on this co-orbital candidate. 

Finally, we performed a joint analysis including the radial velocity and light curve datasets assuming the two-planet scenario to derive precise properties of the known transiting planets. We used the largest evidence model from the radial velocity-only analysis, corresponding to the two-planet case with circular orbits for both planets (this is favored by the Bayesian evidence with a Bayes factor larger than $\mathcal{B}>68$ compared to the eccentric case). The results of this analysis show slightly less massive and smaller planets than previously reported values. For \lhsb,{} we find a planet mass and radius of $m_b=6.48\pm0.46$~\Mearth{} and $R_c=1.641\pm0.048$~\rearth{}, resulting in a bulk density of $\rho_c=7.82^{+0.98}_{-0.88}$~\gcm3. For \lhsc{}, we obtain $m_b=1.77\pm0.17$~\Mearth{} and $R_c=1.185\pm0.044$~\rearth{}, giving $\rho_c=5.81^{+0.87}_{-0.77}$~\gcm3. This analysis provides unprecedented mass and density precisions for such small rocky planets (9\% and 7\% for the outer and inner components, respectively). With these new estimates, both planets lie exactly on the Earth-like bulk density line in the mass-radius diagram (see Fig.~\ref{fig:mr}). Taking advantage of the precise mass and radius measurements, we performed an internal structure analysis considering liquid water conditions and derived a core mass fraction for \lhsb{} and \lhsc{} of $49\pm7$\% and $45\pm10$\%, respectively. It is remarkable that these values are close to the Earth core mass fraction (32\%). Additionally, we found a remarkable water content on \lhsb{}, with the water mass fraction peaking at 4\% (80 times more than on Earth). The posterior distribution for this water mass fraction is truncated at zero, however, thus not allowing a definitive confirmation of a water layer on this planet, but instead providing strong indications that \lhsb{} is a large true water world. This water mass fraction would imply a huge ocean with a depth of $779 \pm 650$~km according to our internal structure model. On the other hand, if water in supercritical phase is considered for \lhsc{}, its maximum WMF is estimated to be 0.06\%, which shows that planet c is likely to be dry or very poor in water. Thus we can conclude that the system of \lhs{} presents a gradient of water mass fraction with irradiation, with an inner dry planet and a wet outer planet. \lhsb{} and \lhsc{} might have acquired their water content because they formed beyond the snow line and then migrated inward. In the case of planet c, a water-dominated atmosphere could have undergone atmospheric escape due to the high irradiation it received from its host star, which caused it to loose most of its water content. {The analysis performed in Sect.~\ref{sec:additional}, however, rejects the presence of additional planets more massive than 1~\Mearth{} in between \lhsc{} and \lhsb{} that could form a resonant chain that could explain the inward migration of these planets from orbits beyond the snow line \citep{delisle17}. Although several mechanisms can explain the disruption of these chains with hot super-Earths \citep{cossou14}, additional theoretical studies should focus on reconciling the interior structure of these planets (requiring that they formed beyond the snow line) with the fact that the planets in the system are currently not in resonance (usual outcome of inward migration in multiplanet systems, see, e.g., \citealt{delisle17}), and the system hosts no additional components in orbits between the two known planets that can create a resonant chain (see Sect.~\ref{sec:additional}).}

With our unprecedentedly precise mass and density measurements for the two transiting rocky worlds, LHS 1140 becomes one of the prime systems for astrobiological studies. The habitable-zone planet \lhsb{} is now known to contain a significant amount of water with a relatively high probability. Previous works have studied the actual state of water layers on the surface of this planet by analyzing the possible eyeball, lobster, or snowball scenarios \citep{yang20}. Future atmospheric studies with the James Webb Space Telescope and extremely large telescopes will be able to distinguish among these scenarios. These observations will greatly benefit from the precise measurements presented here. Added to this, the possible presence of a third rocky planet in the system that may be transiting the host star at periods beyond the habitable zone makes this system a key target for understanding atmospheric properties of rocky worlds  at different stellar irradiations.

%===================================================
\begin{acknowledgements}
This research has been funded by the Spanish State Research Agency (AEI) Projects No.ESP2017-87676-C5-1-R and No. MDM-2017-0737 Unidad de Excelencia "Mar\'ia de Maeztu"- Centro de Astrobiolog\'ia (INTA-CSIC). This work was supported by FCT - Funda\c{c}\~ao para a Ci\^encia e a Tecnologia through national funds and by FEDER through COMPETE2020 - Programa Operacional Competitividade e Internacionaliza\c{c}\~ao by these grants: UID/FIS/04434/2019; UIDB/04434/2020; UIDP/04434/2020; PTDC/FIS-AST/32113/2017 \& POCI-01-0145-FEDER-032113; PTDC/FIS-AST/28953/2017 \& POCI-01-0145-FEDER-028953. N. A.-D. acknowledges the support of FONDECYT project 3180063. XB and JMA acknowledge funding from the European Research Council under the ERC Grant Agreement n. 337591-ExTrA. A.C. acknowledges support by CFisUC projects (UIDB/04564/2020 and UIDP/04564/2020), ENGAGE SKA (POCI-01-0145-FEDER-022217), and PHOBOS (POCI-01-0145-FEDER-029932), funded by COMPETE 2020 and FCT, Portugal. J.P.F. is supported in the form of a work contract funded by national funds through FCT with reference DL57/2016/CP1364/CT0005.
\end{acknowledgements}


%-------------------------------------------------------------------

\bibliographystyle{aa} % style aa.bst
\bibliography{../../biblio2} % your references Yourfile.bib

\appendix

\section{Figures}

\begin{figure*}
\includegraphics[width=1\textwidth]{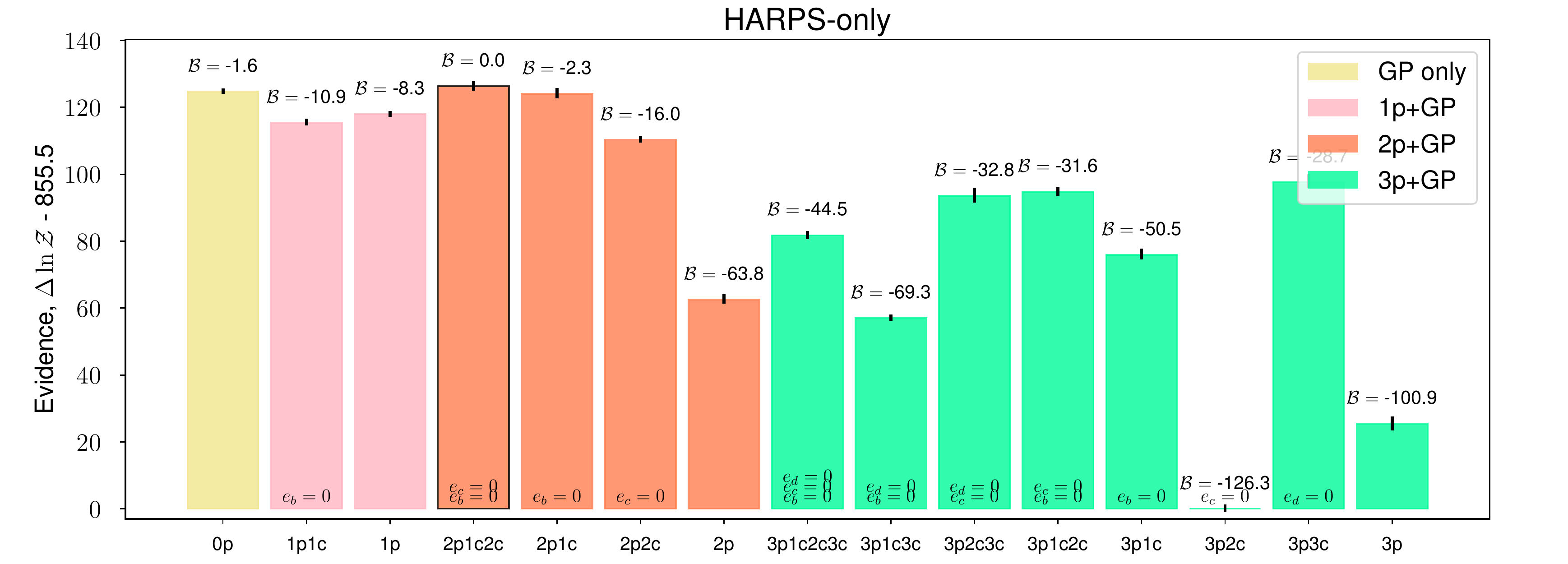}
\includegraphics[width=1\textwidth]{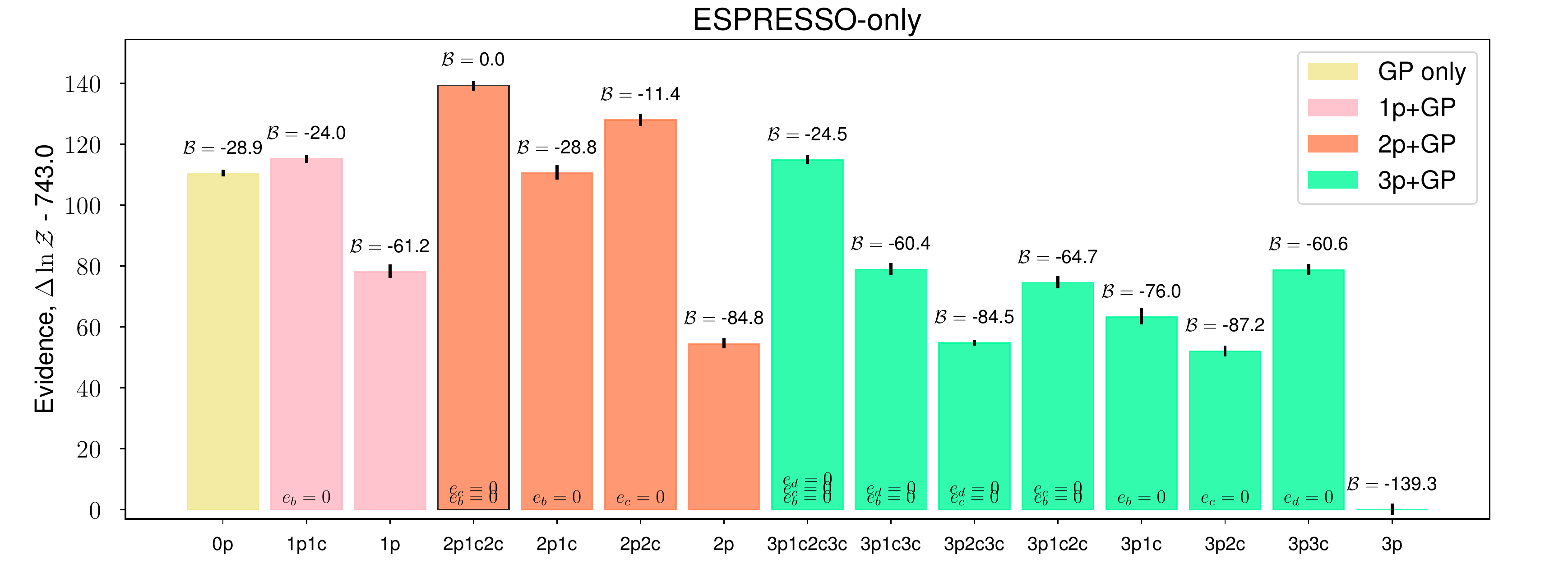}
\includegraphics[width=1\textwidth]{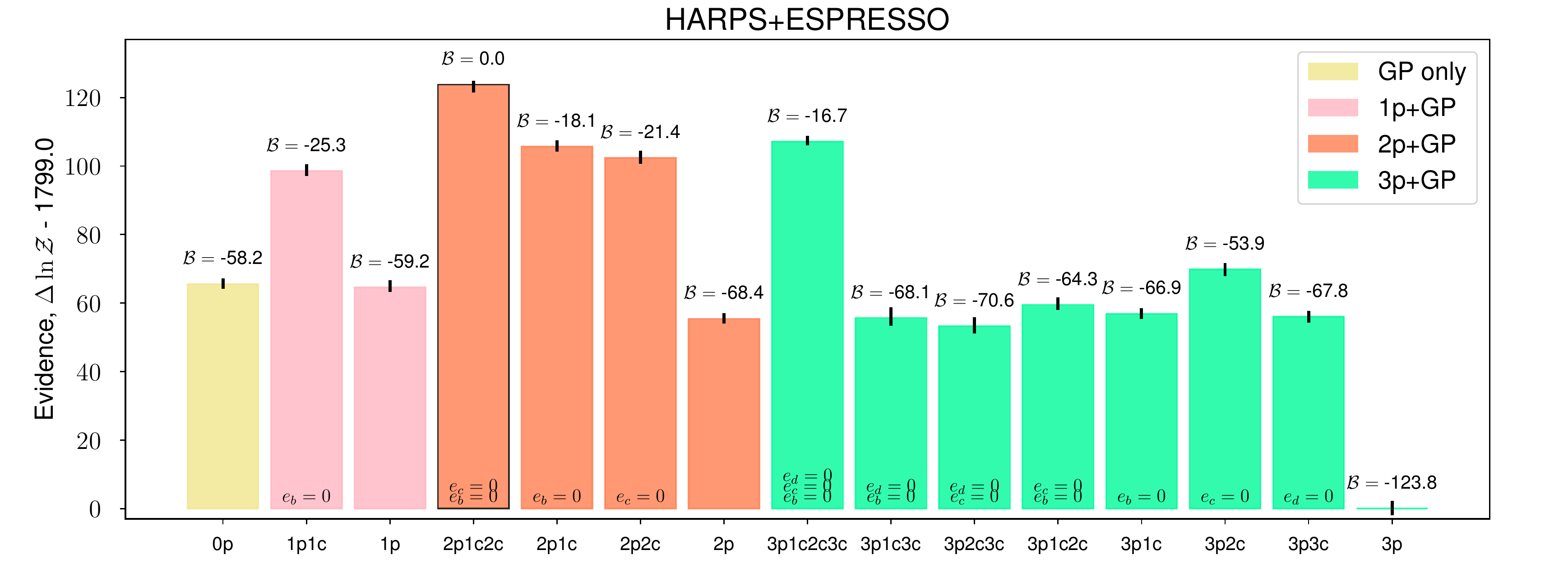}
\caption{Bayesian evidence of different models and datasets (from top to bottom: HARPS-only, ESPRESSO-only, and HARPS+ESPRESSO). Each panel contains 15 different models that include a different number of planets: no planets (GP-only, yellow), one-planet models (pink), two-planet models (red), and three-planet model (green). Each has different orbital configuration assumptions (planets with either circular or eccentric orbits). As an example, model \textit{3p1c2c} corresponds to a three-planet model in which planets 1 (\lhsb{}) and 2 (\lhsc{}) have assumed circular orbits, but planet 3 (\lhsd{}) has free eccentricity. The bars corresponding to each model include this information at the bottom of the bar. At the top of each bar we show the Bayes factor of each model compared to the stronggest evidence model for each dataset. The highest bar corresponds to the strongest evidence model. }
\label{fig:evidence}
\end{figure*}

\section{Tables}

\onecolumn

%%%%%%%%%%%%%%%%%%%%%%%%%%%%%%%%%%%%%%%%%%%%%%%%%%%%%%%%%%

\begin{table}[]
\small
%\centering
\setlength{\extrarowheight}{3pt}
\caption{\label{tab:espresso} Extracted radial velocity, activity indicators, and spectrum properties for the 113 ESPRESSO data points. } % BJD-2458000
\begin{tabular}{lccccccc}

\hline\hline
BJD (days) & RV (km/s) & FWHM (km/s) & Contrast & Asymetry (km/s) & BIS (km/s) & S/N & T$_{\rm exp}$ \\
\hline
\setlength{\extrarowheight}{3pt}
\\

416.71165597 & $-13.2151 \pm 0.0010$ & $3.8791 \pm 0.0020$ & $40.508 \pm 0.021$ & $-0.0255 \pm 0.0012$ & $-0.3698 \pm 0.0020$ & 57.3 & 1820 \\
424.57675194 & $-13.21305 \pm 0.00094$ & $3.8739 \pm 0.0019$ & $41.499 \pm 0.020$ & $-0.0272 \pm 0.0011$ & $5.5868 \pm 0.0019$ & 59.3 & 1820 \\
425.52880438 & $-13.21104 \pm 0.00087$ & $3.8717 \pm 0.0017$ & $41.643 \pm 0.019$ & $-0.0264 \pm 0.0010$ & $-0.4196 \pm 0.0017$ & 62.3 & 1820 \\
431.52720455 & $-13.2165 \pm 0.0013$ & $3.8746 \pm 0.0027$ & $40.584 \pm 0.028$ & $-0.0260 \pm 0.0015$ & $5.5785 \pm 0.0027$ & 46.0 & 1820 \\
431.71439198 & $-13.2155 \pm 0.0012$ & $3.8683 \pm 0.0025$ & $40.896 \pm 0.026$ & $-0.0240 \pm 0.0014$ & $5.5854 \pm 0.0025$ & 48.9 & 1820 \\
432.73355929 & $-13.2142 \pm 0.0011$ & $3.8740 \pm 0.0022$ & $41.134 \pm 0.023$ & $-0.0260 \pm 0.0013$ & $5.5818 \pm 0.0022$ & 53.9 & 1820 \\
434.55594621 & $-13.21287 \pm 0.00095$ & $3.8734 \pm 0.0019$ & $41.479 \pm 0.020$ & $-0.0275 \pm 0.0011$ & $5.5868 \pm 0.0019$ & 58.2 & 1820 \\
435.52959733 & $-13.21287 \pm 0.00094$ & $3.8745 \pm 0.0019$ & $41.448 \pm 0.020$ & $-0.0265 \pm 0.0011$ & $-0.4128 \pm 0.0019$ & 58.9 & 1820 \\
435.65657814 & $-13.2110 \pm 0.0010$ & $3.8706 \pm 0.0021$ & $41.334 \pm 0.022$ & $-0.0255 \pm 0.0012$ & $16.2157 \pm 0.0021$ & 54.8 & 1820 \\
436.5542236 & $-13.2099 \pm 0.0014$ & $3.8752 \pm 0.0028$ & $40.524 \pm 0.029$ & $-0.0247 \pm 0.0016$ & $-0.4230 \pm 0.0028$ & 44.6 & 1820 \\
438.57411678 & $-13.2110 \pm 0.0011$ & $3.8706 \pm 0.0023$ & $40.985 \pm 0.024$ & $-0.0283 \pm 0.0013$ & $5.5881 \pm 0.0023$ & 51.1 & 1820 \\
444.52369713 & $-13.2060 \pm 0.0012$ & $3.8652 \pm 0.0024$ & $40.084 \pm 0.025$ & $-0.0260 \pm 0.0014$ & $5.5892 \pm 0.0024$ & 48.6 & 1820 \\
445.51789345 & $-13.2049 \pm 0.0010$ & $3.8707 \pm 0.0021$ & $40.468 \pm 0.022$ & $-0.0259 \pm 0.0012$ & $-0.2871 \pm 0.0021$ & 54.6 & 1820 \\
446.59162674 & $-13.2087 \pm 0.0010$ & $3.8681 \pm 0.0020$ & $40.958 \pm 0.021$ & $-0.0268 \pm 0.0012$ & $5.5922 \pm 0.0020$ & 55.9 & 1820 \\
447.53735582 & $-13.2067 \pm 0.0011$ & $3.8700 \pm 0.0021$ & $41.383 \pm 0.023$ & $-0.0269 \pm 0.0012$ & $5.5882 \pm 0.0021$ & 53.3 & 1820 \\
448.54585585 & $-13.2084 \pm 0.0010$ & $3.8698 \pm 0.0020$ & $41.470 \pm 0.022$ & $-0.0262 \pm 0.0012$ & $5.5833 \pm 0.0020$ & 55.5 & 1820 \\
450.52180665 & $-13.21199 \pm 0.00099$ & $3.8740 \pm 0.0020$ & $41.358 \pm 0.021$ & $-0.0262 \pm 0.0012$ & $5.5856 \pm 0.0020$ & 56.2 & 1822 \\
452.5317491 & $-13.2110 \pm 0.0011$ & $3.8669 \pm 0.0023$ & $41.250 \pm 0.024$ & $-0.0253 \pm 0.0013$ & $5.5803 \pm 0.0023$ & 50.9 & 1822 \\
455.53442547 & $-13.21539 \pm 0.00082$ & $3.8691 \pm 0.0016$ & $41.823 \pm 0.018$ & $-0.02543 \pm 0.00098$ & $5.5775 \pm 0.0016$ & 63.6 & 1822 \\
456.6122767 & $-13.2136 \pm 0.0012$ & $3.8675 \pm 0.0024$ & $41.119 \pm 0.026$ & $-0.0264 \pm 0.0014$ & $5.5803 \pm 0.0024$ & 48.7 & 1822 \\
458.59472649 & $-5.5007 \pm 0.0026$ & $3.7314 \pm 0.0053$ & $6.8165 \pm 0.0097$ & $0.0175 \pm 0.0015$ & $-15.6343 \pm 0.0053$ & 12.2 & 829 \\
458.61205321 & $-13.21722 \pm 0.00087$ & $3.8674 \pm 0.0017$ & $41.719 \pm 0.019$ & $-0.0264 \pm 0.0010$ & $5.5932 \pm 0.0017$ & 61.5 & 1822 \\
465.647891 & $-13.2126 \pm 0.0014$ & $3.8642 \pm 0.0027$ & $40.590 \pm 0.029$ & $-0.0258 \pm 0.0016$ & $5.5848 \pm 0.0027$ & 44.9 & 1822 \\
470.537938 & $-13.2141 \pm 0.0011$ & $3.8629 \pm 0.0021$ & $40.571 \pm 0.022$ & $-0.0260 \pm 0.0012$ & $5.5893 \pm 0.0021$ & 52.9 & 1822 \\
471.60491706 & $-13.2136 \pm 0.0012$ & $3.8571 \pm 0.0025$ & $40.189 \pm 0.026$ & $-0.0246 \pm 0.0014$ & $5.5841 \pm 0.0025$ & 48.2 & 1822 \\
474.56027993 & $-13.21831 \pm 0.00095$ & $3.8605 \pm 0.0019$ & $40.989 \pm 0.020$ & $-0.0269 \pm 0.0011$ & $5.5941 \pm 0.0019$ & 57.2 & 1822 \\
475.55326919 & $-13.21641 \pm 0.00095$ & $3.8628 \pm 0.0019$ & $41.008 \pm 0.020$ & $-0.0250 \pm 0.0011$ & $5.5844 \pm 0.0019$ & 57.3 & 1822 \\
478.5270003 & $-13.2171 \pm 0.0014$ & $3.8619 \pm 0.0027$ & $40.255 \pm 0.029$ & $-0.0263 \pm 0.0016$ & $5.5827 \pm 0.0027$ & 44.2 & 1822 \\
478.55034832 & $-13.2187 \pm 0.0012$ & $3.8637 \pm 0.0023$ & $41.031 \pm 0.025$ & $-0.0244 \pm 0.0014$ & $-0.3756 \pm 0.0023$ & 49.7 & 1822 \\
479.54501723 & $-13.2173 \pm 0.0016$ & $3.8657 \pm 0.0031$ & $40.247 \pm 0.032$ & $-0.0251 \pm 0.0018$ & $5.5901 \pm 0.0031$ & 40.2 & 1822 \\
479.56803446 & $-13.2219 \pm 0.0013$ & $3.8638 \pm 0.0025$ & $40.809 \pm 0.027$ & $-0.0249 \pm 0.0015$ & $5.5899 \pm 0.0025$ & 46.9 & 1822 \\
480.54341302 & $-13.2221 \pm 0.0010$ & $3.8707 \pm 0.0020$ & $41.320 \pm 0.021$ & $-0.0269 \pm 0.0012$ & $5.5859 \pm 0.0020$ & 55.3 & 1822 \\
487.54308518 & $-13.2163 \pm 0.0013$ & $3.8748 \pm 0.0025$ & $40.774 \pm 0.027$ & $-0.0269 \pm 0.0015$ & $5.5871 \pm 0.0025$ & 47.3 & 1822 \\
489.5446165 & $-13.2089 \pm 0.0012$ & $3.8710 \pm 0.0025$ & $40.906 \pm 0.026$ & $-0.0250 \pm 0.0014$ & $5.5804 \pm 0.0025$ & 48.0 & 1822 \\
494.55238796 & $-13.2080 \pm 0.0018$ & $3.8666 \pm 0.0036$ & $39.344 \pm 0.037$ & $-0.0255 \pm 0.0020$ & $5.5792 \pm 0.0036$ & 36.3 & 2002 \\
496.54962153 & $-13.2107 \pm 0.0014$ & $3.8725 \pm 0.0027$ & $40.302 \pm 0.028$ & $-0.0253 \pm 0.0015$ & $5.5875 \pm 0.0027$ & 45.4 & 1960 \\
499.54805396 & $-13.2157 \pm 0.0011$ & $3.8747 \pm 0.0023$ & $40.762 \pm 0.024$ & $-0.0262 \pm 0.0013$ & $16.1946 \pm 0.0023$ & 50.8 & 1822 \\
500.54867668 & $-13.2116 \pm 0.0012$ & $3.8739 \pm 0.0025$ & $40.627 \pm 0.026$ & $-0.0249 \pm 0.0014$ & $5.5781 \pm 0.0025$ & 48.0 & 1822 \\
508.53278438 & $-13.2143 \pm 0.0011$ & $3.8749 \pm 0.0022$ & $41.263 \pm 0.023$ & $-0.0263 \pm 0.0013$ & $5.5918 \pm 0.0022$ & 52.1 & 1822 \\
510.52924627 & $-13.2087 \pm 0.0012$ & $3.8731 \pm 0.0025$ & $40.829 \pm 0.026$ & $-0.0263 \pm 0.0014$ & $5.5843 \pm 0.0025$ & 48.5 & 1820 \\
511.52220014 & $-13.2043 \pm 0.0019$ & $3.8664 \pm 0.0038$ & $38.862 \pm 0.038$ & $-0.0270 \pm 0.0021$ & $5.5993 \pm 0.0038$ & 35.7 & 1820 \\
517.52482744 & $-13.1919 \pm 0.0033$ & $3.8373 \pm 0.0066$ & $36.786 \pm 0.063$ & $-0.0178 \pm 0.0033$ & $5.5849 \pm 0.0066$ & 24.2 & 1820 \\
668.7930973 & $-13.2075 \pm 0.0014$ & $3.8774 \pm 0.0028$ & $40.422 \pm 0.029$ & $-0.0243 \pm 0.0016$ & $5.5887 \pm 0.0028$ & 43.8 & 1916 \\
670.84860195 & $-13.20980 \pm 0.00098$ & $3.8783 \pm 0.0020$ & $41.320 \pm 0.021$ & $-0.0270 \pm 0.0011$ & $16.0227 \pm 0.0020$ & 56.9 & 1916 \\
673.82845489 & $-13.21230 \pm 0.00094$ & $3.8833 \pm 0.0019$ & $41.776 \pm 0.020$ & $-0.0257 \pm 0.0011$ & $-0.4181 \pm 0.0019$ & 57.4 & 1916 \\
674.90806209 & $-13.21140 \pm 0.00084$ & $3.8765 \pm 0.0017$ & $41.492 \pm 0.018$ & $-0.02568 \pm 0.00099$ & $16.0481 \pm 0.0017$ & 63.2 & 1916 \\
... &&&&&&&\\
\setlength{\extrarowheight}{3pt}
\\
\hline  

\end{tabular}
\end{table}

%%%%%%%%%%%%%%%%%%%%%%%%%%%%%%%%%%%%%%%%%%%%%%%%%%%%%%%%%%

\newpage
\setlength{\extrarowheight}{3pt}
\begin{longtable}{lccc}
\caption{\label{tab:rv3p}Priors and posterior distributions for the radial velocity analysis (see Sect.~\ref{sec:rv3p}).}\\
\hline\hline
Parameter & Prior & 2p+GP & 3p+GP \\ 
\hline
\endfirsthead
\multicolumn{4}{l}{{\bfseries \tablename\ \thetable{} -- continued from previous page}} \\
\hline
Parameter & Prior & 2p+GP & 3p+GP \\ 
\hline
\endhead
\multicolumn{4}{l}{{Continued on next page}} \\ 
\hline
\endfoot
\hline
\multicolumn{4}{l}{Notes:}\\
\multicolumn{4}{l}{$\bullet$ $\mathcal{N}(\mu,\sigma^{2})$: Normal distribution with mean $\mu$ and width $\sigma^{2}$}\\
\multicolumn{4}{l}{$\bullet$ $\mathcal{U}(a,b)$: Uniform distribution between $a$ and $b$}\\
\multicolumn{4}{l}{$\bullet$ $\mathcal{LU}(a,b)$: Log-uniform distribution between $a$ and $b$}\\
\multicolumn{4}{l}{$\bullet$ $\mathcal{T}(\mu,\sigma^{2},a,b)$: Truncated normal distribution with mean $\mu$ and width $\sigma^{2}$, between $a$ and $b$}\\
\endlastfoot
\\
\multicolumn{3}{l}{\it \lhsb{}}\\
Orbital period, $P_b$ [days]                            & $\mathcal{G}$(24.736959,0.1)           & $24.7379^{+0.0038}_{-0.0038}$ &  $24.7385^{+0.0037}_{-0.0036}$ \\
Time of mid-transit, $T_{\rm 0,b}-2400000$ [days]       & $\mathcal{G}$(56915.71154,0.1)         & $56915.71^{+0.10}_{-0.10}$ & $56915.71^{+0.10}_{-0.10}$ \\
RV semi-amplitude, $K_{\rm b}$ [m/s]                    & $\mathcal{U}$(0.0,20.0)                & $4.21^{+0.24}_{-0.24}$ & $4.12^{+0.23}_{-0.23}$ \\
Orbital eccentricity, $e_{\rm b}$ & Fixed (0.0) & $<0.096$ &  $<0.086$\\
Arg. periastron, $\omega_{\rm b}$ [deg.] & Fixed ($90^{\circ}$) & & \\
Planet mass, $m_{b}\sin{i_b}$ [\Mearth{}]               & (derived)                              & $6.09^{+0.48}_{-0.47}$ & $5.96^{+0.46}_{-0.45}$ \\

\\
\multicolumn{3}{l}{\it \lhsc{}}\\
Orbital period, $P_c$ [days]                            & $\mathcal{G}$(3.777931,0.1)            & $3.77728^{+0.00046}_{-0.00045}$ & $3.77726^{+0.00046}_{-0.00045}$ \\
Time of mid-transit, $T_{\rm 0,c}-2400000$ [days]       & $\mathcal{G}$(58226.843169,0.1)        & $58226.814^{+0.052}_{-0.051}$ & $58226.812^{+0.052}_{-0.051}$ \\
RV semi-amplitude, $K_{\rm c}$ [m/s]                    & $\mathcal{U}$(0.0,10.0)                & $2.22^{+0.20}_{-0.20}$ & $2.20^{+0.20}_{-0.20}$ \\
Orbital eccentricity, $e_{\rm c}$ & Fixed (0.0)  & $<0.274$ & $<0.252$\\
Arg. periastron, $\omega_{\rm c}$ [deg.] & Fixed ($90^{\circ}$) & & \\
Planet mass, $m_{c}\sin{i_c}$ [\Mearth{}]               & (derived)                              & $1.71^{+0.18}_{-0.18}$ & $1.70^{+0.18}_{-0.18}$ \\

\\
\multicolumn{3}{l}{\it \lhsd{} (candidate)}\\
Orbital period, $P_d$ [days]                            & $\mathcal{U}$(70.0,120.0)              &  & $79.22^{+0.55}_{-0.58}$ \\
Time of mid-transit, $T_{\rm 0,d}-2400000$ [days]       & $\mathcal{U}$(58350.0,58400.0)         &  & $58382.3^{+3.3}_{-3.3}$ \\
RV semi-amplitude, $K_{\rm d}$ [m/s]                    & $\mathcal{U}$(0.0,10.0)                &  & $2.21^{+0.59}_{-0.57}$ \\
Orbital eccentricity, $e_{\rm d}$ & Fixed (0.0)  &  & $<0.45$\\
Arg. periastron, $\omega_{\rm c}$ [deg.] & Fixed ($90^{\circ}$) & & \\
Planet mass, $m_{d}\sin{i_d}$ [\Mearth{}]               & (derived)                              &  & $4.8^{+1.3}_{-1.2}$ \\                                                 
\\
\multicolumn{3}{l}{\it Instrument parameters}\\
$\delta_{\rm ESPRESSOpre}$ [m/s]                        & $\mathcal{U}$(-15.0,-10.0)             & $-13212.3^{+2.0}_{-2.0}$ & $-13212.2^{+2.1}_{-1.9}$ \\
$\delta_{\rm ESPRESSOpost}$ [m/s]                       & $\mathcal{U}$(-15.0,-10.0)             & $-13210.7^{+1.8}_{-1.9}$ & $-13210.6^{+1.9}_{-1.9}$ \\
$\delta_{\rm HARPSc}$ [m/s]                             & $\mathcal{U}$(-15.0,-10.0)             & $-13238.6^{+1.2}_{-1.3}$ & $-13238.9^{+1.2}_{-1.3}$ \\
$\sigma_{\rm ESPRESSOpre}$ [m/s]                        & $\mathcal{U}$(0.0,0.005)               & $1.35^{+0.34}_{-0.32}$ & $1.29^{+0.33}_{-0.30}$ \\
$\sigma_{\rm ESPRESSOpost}$ [m/s]                       & $\mathcal{U}$(0.0,0.005)               & $1.06^{+0.21}_{-0.19}$ & $1.10^{+0.21}_{-0.20}$ \\
$\sigma_{\rm HARPSc}$ [m/s]                             & $\mathcal{U}$(0.0,0.005)               & $0.42^{+0.43}_{-0.29}$ & $0.40^{+0.42}_{-0.28}$ \\

$\delta_{\rm FWHM,ESPRESSOpre}$ [km/s]                  & $\mathcal{G}$(3.8,0.1)                 & $3.8713^{+0.0062}_{-0.0063}$ & $3.8714^{+0.0074}_{-0.0073}$ \\
$\delta_{\rm FWHM,ESPRESSOpost}$ [km/s]                 & $\mathcal{G}$(3.8,0.1)                 & $3.8778^{+0.0063}_{-0.0063}$ & $3.8785^{+0.0076}_{-0.0070}$ \\
$\delta_{\rm FWHM,HARPSc}$ [km/s]                       & $\mathcal{G}$(3.0,0.1)                 & $2.9737^{+0.0039}_{-0.0042}$ & $2.9730^{+0.0046}_{-0.0050}$ \\
$\sigma_{\rm FWHM,ESPRESSOpre}$ [m/s]                   & $\mathcal{U}$(0.0,0.01)                & $1.22^{+0.74}_{-0.73}$ & $1.34^{+0.71}_{-0.77}$ \\
$\sigma_{\rm FWHM,ESPRESSOpost}$ [m/s]                  & $\mathcal{U}$(0.0,0.01)                & $3.06^{+0.52}_{-0.46}$ & $3.06^{+0.52}_{-0.48}$ \\
$\sigma_{\rm FWHM,HARPSc}$ [m/s]                        & $\mathcal{U}$(0.0,0.01)                & $1.9^{+1.4}_{-1.3}$ & $2.2^{+1.3}_{-1.4}$ \\

\\
\multicolumn{3}{l}{\it GP hyperparameters}\\
$\eta_{\rm 1,FWHM}$ [m/s]                               & $\mathcal{U}$(-6.0,6.0)                & $2.39^{+0.16}_{-0.15}$ & $2.46^{+0.19}_{-0.17}$ \\
$\eta_1$ [m/s]                                          & $\mathcal{LU}$(-5.0,5.0)               & $1.21^{+0.21}_{-0.18}$ & $1.10^{+0.24}_{-0.20}$ \\
$\eta_2$ [days]                                         & $\mathcal{U}$(100.0,500.0)             & $133^{+40}_{-23}$ & $135^{+38}_{-22}$ \\
$\eta_3$ [days]                                         & $\mathcal{G}$(131.0,5.0)               & $132.0^{+2.5}_{-3.0}$ & $130.9^{+2.7}_{-3.7}$ \\
$\eta_4$                                                & $\mathcal{LU}$(-2.0,2.0)               & $-0.97^{+0.20}_{-0.18}$ & $-0.80^{+0.25}_{-0.24}$ \\
\\
\hline 
 \\
 $\ln{\mathcal{Z}}$  & & $1922.8^{+0.7}_{-1.8}$& $1906.1^{+1.3}_{-0.5}$ \\
 \\
 \hline
 
% \end{tabular}
%\tablefoot{
%{Notes:}\\
%{$\bullet$ $\mathcal{N}(\mu,\sigma^{2})$: Normal distribution with mean $\mu$ and width $\sigma^{2}$}\\
%{$\bullet$ $\mathcal{U}(a,b)$: Uniform distribution between $a$ and $b$}\\
%{$\bullet$ $\mathcal{LU}(a,b)$: Log-uniform distribution between $a$ and $b$}\\
%{$\bullet$ $\mathcal{T}(\mu,\sigma^{2},a,b)$: Truncated normal distribution with mean $\mu$ and width $\sigma^{2}$, between $a$ and $b$}\\
%}

\end{longtable}

%%%%%%%%%%%%%%%%%%%%%%%%%%%%%%%%%%%%%%%%%%%%%%%%%%%%%%%%%%
\newpage
\setlength{\extrarowheight}{3pt}
\begin{longtable}{lcc}
\caption{\label{tab:joint}Priors and posterior distributions for the parameters modeled in the joint fit analysis (see Sect.~\ref{sec:joint}).}\\
\hline\hline
{Planet parameters} & Prior & Posterior \\ 
\hline
\endfirsthead
\caption{continued.}\\
\hline\hline
{Planet parameters} & Prior & Posterior \\ 
\hline
\endhead
\hline
\endfoot
{Notes:}\\
\multicolumn{3}{l}{$\bullet$ $\mathcal{N}(\mu,\sigma^{2})$: Normal distribution with mean $\mu$ and width $\sigma^{2}$}\\
\multicolumn{3}{l}{$\bullet$ $\mathcal{U}(a,b)$: Uniform distribution between $a$ and $b$}\\
\multicolumn{3}{l}{$\bullet$ $\mathcal{LU}(a,b)$: Log-uniform distribution between $a$ and $b$}\\
\multicolumn{3}{l}{$\bullet$ $\mathcal{T}(\mu,\sigma^{2},a,b)$: Truncated normal distribution with mean $\mu$ and width $\sigma^{2}$, between $a$ and $b$}\\
\endlastfoot
\setlength{\extrarowheight}{3pt}
\\
\multicolumn{3}{l}{\it Stellar Parameters}\\
Stellar radius, $R_{\star}$ [$R_{\odot}$] & $\mathcal{T}$(0.2139,0.0041,0,1) & $0.2134^{+0.0036}_{-0.0034}$ \\
Stellar mass, $M_{\star}$ [$M_{\odot}$] & $\mathcal{T}$(0.179,0.014,0.,1) & $0.191^{+0.012}_{-0.011}$ \\
Limb darkening coefficient, $u_1$ & $\mathcal{T}$(0.1858804,0.1,0.,1) & $0.231^{+0.091}_{-0.092}$ \\
Limb darkening coefficient, $u_2$ & $\mathcal{T}$(0.49001512,0.1,0,1) & $0.517^{+0.095}_{-0.095}$ \\
Stellar luminosity, $L_{\star}$ [$L_{\odot}$] & (derived) & $0.477^{+0.022}_{-0.021}$ \\

\\
\multicolumn{3}{l}{\it \lhsb{}}\\
Orbital period, $P_b$ [days] & $\mathcal{G}$(24.736959,0.0004) & $24.73694^{+0.00041}_{-0.00040}$ \\
Time of mid-transit, $T_{\rm 0,b}-2400000$ [days] & $\mathcal{U}$(58399.0,58401.0) & $58399.9303^{+0.0012}_{-0.0013}$ \\
Planet mass, $M_b$ [$M_{\oplus}$] & $\mathcal{U}$(0.0,50.0) & $6.38^{+0.46}_{-0.44}$ \\
Planet radius, $R_b$ [$R_{\oplus}$] & $\mathcal{T}$(1.727,0.1,0,10) & $1.635^{+0.046}_{-0.046}$ \\
Orbital inclination, $i_{\rm b}$ [deg.] & $\mathcal{T}$(89.89,0.05,70,90) & $89.877^{+0.049}_{-0.045}$ \\
Planet density, $\rho_{b}$ [$g\cdot cm^{-3}$] & (derived) & $8.04^{+0.84}_{-0.80}$ \\
Transit depth, $\Delta_{b}$ [ppt] & (derived) & $4.93^{+0.27}_{-0.27}$ \\
Orbit semi-major axis, $a_{b}$ [AU] & (derived) & $0.0957^{+0.0019}_{-0.0019}$ \\
Relative orbital separation, $a_{b}/R_{\star}$ & (derived) & $96.4^{+2.2}_{-2.1}$ \\
Transit duration, $T_{\rm 14,b}$ [hours] & (derived) & $2.055^{+0.048}_{-0.049}$ \\
Planet surface gravity, $g_{\rm b}$ [$m \cdot s^{-2}$] & (derived) & $23.4^{+1.9}_{-2.0}$ \\
Incident Flux, $F_{\rm inc,b}$ [$F_{{\rm inc},\oplus}$] & (derived) & $4.98^{+0.23}_{-0.22}$ \\
Stellar effective incident flux, $S_{b}$ [$S_{\oplus}$] & (derived) & $0.477^{+0.022}_{-0.021}$ \\
Stellar luminosity, $L_{\star}$ [$L_{\odot}$] & (derived) & $0.477^{+0.022}_{-0.021}$ \\
Equilibrium temperature, $T_{\rm eq,b}$ [K] & (derived) & $378.9^{+4.3}_{-4.2}$ \\

\\
\multicolumn{3}{l}{\it \lhsc{}}\\
Orbital period, $P_c$ [days] & $\mathcal{G}$(3.777931,3e-05) & $3.777929^{+0.000030}_{-0.000030}$ \\
Time of mid-transit, $T_{\rm 0,c}-2400000$ [days] & $\mathcal{G}$(58389.2939,0.1) & $58389.29382^{+0.00081}_{-0.00082}$ \\
Planet mass, $M_c$ [$M_{\oplus}$] & $\mathcal{U}$(0.0,50.0) & $1.76^{+0.17}_{-0.16}$ \\
Orbital inclination, $i_{\rm c}$ [deg.] & $\mathcal{T}$(89.92,0.05,70,90) & $89.913^{+0.046}_{-0.049}$ \\
Planet radius, $R_c$ [$R_{\oplus}$] & $\mathcal{T}$(1.282,0.1,0,10) & $1.169^{+0.037}_{-0.038}$ \\
Planet density, $\rho_{c}$ [$g\cdot cm^{-3}$] & (derived) & $6.07^{+0.81}_{-0.74}$ \\
Transit depth, $\Delta_{c}$ [ppt] & (derived) & $2.52^{+0.16}_{-0.15}$ \\
Orbit semi-major axis, $a_{c}$ [AU] & (derived) & $0.02734^{+0.00054}_{-0.00054}$ \\
Relative orbital separation, $a_{c}/R_{\star}$ & (derived) & $27.53^{+0.62}_{-0.61}$ \\
Transit duration, $T_{\rm 14,c}$ [hours] & (derived) & $1.100^{+0.025}_{-0.024}$ \\
Planet surface gravity, $g_{\rm c}$ [$m \cdot s^{-2}$] & (derived) & $12.6^{+1.4}_{-1.3}$ \\
Incident Flux, $F_{\rm inc,c}$ [$F_{{\rm inc},\oplus}$] & (derived) & $61.0^{+2.8}_{-2.6}$ \\
Stellar effective incident flux, $S_{c}$ [$S_{\oplus}$] & (derived) & $5.85^{+0.27}_{-0.25}$ \\
Equilibrium temperature, $T_{\rm eq,c}$ [K] & (derived) & $708.9^{+8.0}_{-7.8}$ \\

\\
\multicolumn{3}{l}{\it Instrument parameters}\\
LC level & $\mathcal{U}$(-500.0,500.0) & $-54^{+19}_{-19}$ \\
Dilution factor & $\mathcal{T}$(0.052,0.001,0.,1.) & $0.0520^{+0.0010}_{-0.0010}$ \\
LC jitter [ppm] & $\mathcal{U}$(0.0,2000.0) & $50^{+52}_{-35}$ \\
$\delta_{\rm ESPRESSOpre}$ [m/s] & $\mathcal{U}$(-15.0,-10.0) & $-13.2124^{+0.0018}_{-0.0018}$ \\
$\delta_{\rm ESPRESSOpost}$ [m/s] & $\mathcal{U}$(-15.0,-10.0) & $-13.2107^{+0.001.8}_{-0.0018}$ \\
$\delta_{\rm HARPSc}$ [m/s] & $\mathcal{U}$(-15.0,-10.0) & $-13.2386^{+0.0012}_{-0.0011}$ \\
$\sigma_{\rm ESPRESSOpre}$ [m/s] & $\mathcal{U}$(0.0,0.005) & $1.32^{+0.34}_{-0.32}$ \\
$\sigma_{\rm ESPRESSOpost}$ [m/s] & $\mathcal{U}$(0.0,0.005) & $1.07^{+0.20}_{-0.19}$ \\
$\sigma_{\rm HARPSc}$ [m/s] & $\mathcal{U}$(0.0,0.005) & $0.41^{+0.40}_{-0.29}$ \\
\\
\multicolumn{3}{l}{\it GP hyperparameters}\\
$\eta_{\rm 1,FWHM}$ [m/s] & $\mathcal{U}$(-6.0,6.0) & $2.40^{+0.15}_{-0.13}$ \\
$\eta_1$ [m/s] & $\mathcal{LU}$(-5.0,5.0) & $1.16^{+0.18}_{-0.16}$ \\
$\eta_2$ [days] & $\mathcal{U}$(100.0,500.0) & $135^{+39}_{-24}$ \\
$\eta_3$ [days] & $\mathcal{G}$(131.0,5.0) & $131.4^{+2.3}_{-2.9}$ \\
$\eta_4$ & $\mathcal{LU}$(-2.0,2.0) & $-1.04^{+0.16}_{-0.15}$ \\

\hline

\end{longtable}

%%%%%%%%%%%%%%%%%%%%%%%%%%%%%%%%%%%%%%%%%%%%%%%%%%%%%%%%%%

\end{document}